\documentclass{paper}

\usepackage[T1]{fontenc}
\usepackage{lmodern}

\newlength{\verbindent}
\usepackage{fancyvrb}

\usepackage{ctable}

\usepackage{fmslash}
\newcommand{\pslash}{\fmslash[1mu]{p}}

\newcommand{\var}[1]{\texttt{\textsl{#1}}}
\newcommand{\out}{\mbox{$\Rightarrow{}$}}
\newcommand{\noout}{\phantom{$\Rightarrow{}$}}
\newcommand{\lb}{\texttt{\{}}
\newcommand{\rb}{\texttt{\}}}

\preprintnumber{IPPP/15/42\\DCPT/15/84}

\newcommand{\code}[1]{\texttt{#1}}

\title{HEPMath 1.4: A Mathematica Package for Semi-Automatic Computations in
High Energy Physics}

\author{%
  Martin Wiebusch\email{martin.wiebusch@durham.ac.uk}%
}

\institute{Institute for Particle Physics Phenomenology,
  Department of Physics, Durham University, Durham CH1 3LE, United Kingdom}%

\abstract{This article introduces the Mathematica package \emph{HEPMath} which
  provides a number of utilities and algorithms for High Energy Physics
  computations in Mathematica. Its functionality is similar to packages like
  FormCalc or FeynCalc, but it takes a more complete and extensible approach to
  implementing common High Energy Physics notations in the Mathematica language,
  in particular those related to tensors and index contractions. It also
  provides a more flexible method for the generation of numerical code which is
  based on new features for C code generation in Mathematica. In particular it
  can automatically generate \emph{Python extension modules} which make the
  compiled functions callable from Python, thus eliminating the need to write
  any code in a low-level language like C or Fortran. It also contains seamless
  interfaces to LHAPDF, FeynArts, and LoopTools.}

\begin{document}
\maketitlepage

\tableofcontents

%%%%%%%%%%%%%%%%%%%%%%%%%%%%%%%%%%%%%%%%%%%%%%%%%%%%%%%%%%%%%%%%%%%%%%%%%%%%%%%%
\section{Introduction}
%%%%%%%%%%%%%%%%%%%%%%%%%%%%%%%%%%%%%%%%%%%%%%%%%%%%%%%%%%%%%%%%%%%%%%%%%%%%%%%%

In many areas of High Energy Physics the scientific progress is intrinsically
linked to progress in the development of computational tools and the
implementation of computational algorithms. A prime example for this interplay
is the numerical computation and integration of scattering cross sections.  For
tree-level cross sections in the Standard Model (SM) automatisation has been
accomplished about 10 years ago \cite{hep-ph/9401258, hep-ph/9807565,
  hep-ph/0007053, hep-ph/0002082, hep-ph/0102195, hep-ph/0109036,
  hep-ph/0403113, arXiv:0710.2427, arXiv:0708.4233, arXiv:0706.2334,
  arXiv:0811.4622, arXiv:1207.6082}. The implementation of virtually arbitrary
models has been automated in \cite{Semenov:1997qm, arXiv:0805.0555,
  arXiv:0806.4194, arXiv:0906.2474, arXiv:0909.2863, arXiv:1310.1921,
  arXiv:1309.7223}. More recently substantial progress has been made towards the
automatisation of the computation next-to-leading order cross sections
\cite{arXiv:1002.2581, arXiv:1103.0621, arXiv:1111.5206, arXiv:1310.2808,
  arXiv:1405.0301}. These tools aim to be \emph{fully} automatic in the sense
that they implement a fixed computational scheme and the only user input
required is information about the process, model parameters, perturbative order
etc.

There are, however, other types of calculations in high energy physics where
full automatisation (in the above-mentioned sense) is either not possible or not
yet available. Such cases include the computation of Wilson coefficients or
anomalous dimensions in effective field theories and calculations where the
result is needed in analytic form or where process-specific manipulations
(expansions in external momenta etc.) are required at some stage in the symbolic
computation. These computations can still be greatly simplified by
automatisation, but they call for a different type of tool, which I will refer
to as \emph{semi-automatic} tools. Semi-automatic tools try to mirror
conventional mathematical notations in a programming environment and provide a
common interface to as many standard algorithms as possible (e.g.\ the
computation of Dirac traces or the reduction of loop integrals).  To offer the
necessary flexibility they must, however, leave control over the computational
scheme in the hands of the user. The \emph{HEPMath} package presented in this
paper falls in this category. It provides the following functionality:
\begin{enumerate}
  \item Consistent integration of tensors with symbolic indices and index
    contractions in Mathematica, along with functions to manipulate (expand,
    collect etc.) such expressions. Arbitrary tensors and index types can
    be defined by the user.
  \item Implementation of the basic building blocks of Feynman amplitudes, such
    as Dirac spinors, Dirac matrices, Lorentz vectors, colour matrices
    etc.\ and their algebraic properties.
  \item Computation of traces over Dirac matrices in 4 and $D$ dimensions.
  \item Simplification of colour structures.
  \item Passarino-Veltman reduction of one-loop integrals. Currently only the
    reduction to tensor integrals is implemented, but this is likely to change
    in the future.
  \item Seamless interfaces to FeynArts \cite{Comp.Phys.Comm.60.165,
    hep-ph/0012260}, LoopTools \cite{hep-ph/9807565} and LHAPDF \cite{LHAPDF}.
  \item Generation and compilation of numerical code, based on Mathematica's
    \code{Compile} and \code{CCodeGenerate} functions.
  \item Generation of \emph{Python extension modules} which make the generated
    functions callable from Python.
\end{enumerate}

Feature 8 is new to the best of my knowledge, but there are of course several
other publicly available semi-automatic tools which provide a subset of the
above-mentioned functionality. Some also provide functionality that is not
present in HEPMath (yet). In the following I will briefly review these tools
and highlight the differences and similarities with HEPMath.

\begin{description}
\item[FORM] \cite{math-ph/0010025} is clearly one of the most established tools
  for symbolic computations in High Energy Physics. Its functionality covers
  items 1 to 3 and it is highly optimised for handling large polynomial
  expressions.  It is a standalone program with its own input language which,
  despite its flexibility, can not be regarded as a general purpose programming
  language.  The design of FORM puts performance before ease of use, while
  HEPMath is mainly intended as a convenience tool. Simply put, FORM will
  reduce the \emph{run time} of your symbolic computations while HEPMath tries
  to reduce the \emph{development time}. In many applications the latter vastly
  outweighs the former, and HEPMath is written for these cases.
\item[TRACER] \cite{Comp.Phys.Comm.74.265} is a Mathematica package which
  calculates traces over Dirac matrices. Its functionality covers item 3 and
  some aspects of item 2. It can be regarded as an ancestor of HEPMath, since
  the HEPMath algorithm for the computation of Dirac traces was taken from
  TRACER. However, the TRACER syntax for constructing tensor expressions (such
  as contractions between Lorentz vectors etc.) is ambiguous\footnote{Try to
    make a substitution like $p\to xk$ in TRACER, where $p$ and $k$ are Lorentz
    vectors and $x$ a scalar. There is no way to tell TRACER that $x$ is the
    scalar and $k$ the vector.} and was completely re-designed in HEPMath.
\item[FormCalc] \cite{hep-ph/9807565, hep-ph/0210220, hep-ph/0406288,
  hep-ph/0506201, hep-ph/0601248, hep-ph/0607049, hep-ph/0611273} is a
  Mathematica package for the calculation of tree-level and one-loop
  amplitudes. It is developed in parallel with FeynArts
  \cite{Comp.Phys.Comm.60.165, hep-ph/0012260} and LoopTools
  \cite{hep-ph/9807565} and its functionality covers items 2 to 7. It uses FORM
  as a back-end for parts of the symbolic computation and thus offers the
  performance of FORM through a user friendly Mathematica interface. Its
  approach differs from HEPMath in the sense that its computational scheme,
  although configurable with many options, is ultimately pre-defined and out of
  the user's control. Also, the structure of numerical code generated with
  FormCalc is mostly fixed while the code generation mechanism of HEPMath
  offers a lot more freedom.
\item[FeynCalc] \cite{Comp.Phys.Comm.64.345} is a Mathematica package whose
  goals and design principles are very similar to those of HEPMath. Its
  functionality includes items 2 to 6. It also provides several symbolic
  algorithms which are not present in HEPMath. In addition to features 1, 7,
  and 8 an advantage of HEPMath over FeynCalc is the absence of an `internal'
  representation of expressions involving tensors (i.e. Lorentz vectors, Dirac
  matrices etc.), which simplifies the implementation of additional symbolic
  algorithms by the user.
\end{description}

For any but the most trivial calculations it is generally unwise to rely on a
single tool for automatic computations. With the exception of TRACER, HEPMath
has no code overlap with any of the above-mentioned tools. It can therefore be
used as an independent check of automatic calculations performed with any of
them (except TRACER). The goal of the HEPMath project is to surpass these tools
not so much in speed but in flexibility and ease-of-use. But of course such
criteria are somewhat subjective.

This article is a user's guide for the HEPMath package. It documents its main
functionality and illustrates its use with many examples.

%%%%%%%%%%%%%%%%%%%%%%%%%%%%%%%%%%%%%%%%%%%%%%%%%%%%%%%%%%%%%%%%%%%%%%%%%%%%%%%%
\section{Installation}
%%%%%%%%%%%%%%%%%%%%%%%%%%%%%%%%%%%%%%%%%%%%%%%%%%%%%%%%%%%%%%%%%%%%%%%%%%%%%%%%

The HEPMath package was developed with Mathematica 9 on a 64bit Linux platform.
I will only describe the installation process on Linux here. The latest version
of the HEPMath package is available from
\begin{center}
  \href{http://hepmath.hepforge.org}{\texttt{http://hepmath.hepforge.org}}
\end{center}
After downloading the tarball \code{hepmath-}\var{x.y}\code{.tar.gz} (where
`\var{x.y}' stands for the current version number) run the following commands
in a terminal:
\begin{Verbatim}[commandchars=\\\{\}]
tar -zxvf hepmath-\var{x.y}.tar.gz
cd hepmath-\var{x.y}
./configure \var{[options]}
make
make install
\end{Verbatim}
All necessary information for the installation can be specified through command
line options to the \code{configure} script. The order of the arguments is
irrelevant.

As for all GNU packages, the installation prefix for various libraries and
Python modules included in HEPMath can be set with the \code{-{}-prefix}
option. The default setting typically requires administrator rights, and if you
don't have them you might want to change it, e.g.\ to
\begin{Verbatim}[commandchars=\\\{\}]
--prefix=\var{my-prefix}
\end{Verbatim}
If you want to generate Python extension modules with HEPMath you may want to use
a prefix that your Python interpreter searches for Python packages. On my system
this is \code{\$HOME/.local}.

To install only the Mathematica packages without the LHAPDF and LoopTools
interfaces no additional options need to be passed to \code{configure}. Note
that \code{make} needs to execute a few Mathematica scripts. It assumes that the
command which starts the command line version of Mathematica is called
\code{math}. If the executable has a different name, say \code{math9}, on your
system (e.g. because you have several Mathematica versions installed) you can
tell \code{configure} about this by passing
\begin{Verbatim}
MATH=math9
\end{Verbatim}
as a command line argument.  The Mathematica modules will be installed in
\texttt{\$HOME/.Mathematica/Applications/}, independent of your
\code{-{}-prefix} setting.  You can change the installation directory of the
Mathematica modules to \var{path} with
\begin{Verbatim}[commandchars=\\\{\}]
MATHDIR=\var{path}
\end{Verbatim}
However, it is your responsibility to assure that Mathematica searches \var{path}
for packages, e.g. by setting the \code{\$UserBaseDirectory} variable in
Mathematica accordingly.

HEPMath can access parton distribution functions by interfacing with the LHAPDF
library. This obviously requires a working installation of the LHAPDF
package. In particular you must make sure that the library \code{libLHAPDF.so}
from by the LHAPDF package is installed in a directory where your dynamic linker
can find it. This can, for example, be accomplished by adding the directory
containing \code{libLHAPDF.so} to your \code{LD\_LIBRARY\_PATH} environment
variable. The HEPMath interface to LHAPDF was tested with LHAPDF version
6.1.3. To enable it use one of the following options:
\begin{Verbatim}[commandchars=\\\{\}]
--with-lhapdf
--with-lhapdf=\var{path}
\end{Verbatim}
Use the first form if you have installed LHAPDF under a standard prefix
(e.g. \code{/usr} or \code{/usr/local}) or if you have configured your system so
that your C compiler and linker find the LHAPDF headers and libraries without
additional flags. The second form allows you to specify the prefix
\var{path} under which LHAPDF is installed. Alternatively, you can specify the
compiler and linker flags required for LHAPDF with
\begin{Verbatim}[commandchars=\\\{\}]
LHAPDF_CFLAGS=\var{compile-flags}
LHAPDF_LIBS=\var{link-flags}
\end{Verbatim}

One-loop tensor integrals can be evaluated numerically in HEPMath through an
interface to the LoopTools package. The LoopTools interface was tested with
LoopTools version 2.9.  To install it you will most likely need to re-compile
the LoopTools library. The reason is that LoopTools creates a static library by
default, while the HEPMath interface requires a dynamic library. Converting
a static to a dynamic library is only possible when the static library was
compiled with the right flags, and that's why you need to re-compile. On most
Linux platforms this can be done as follows:
\begin{Verbatim}[commandchars=\\\{\}]
cd \var{looptools-path}
make clean
CFLAGS=-fPIC FFLAGS=-fPIC \var{defs} ./configure \var{opts}
make
\end{Verbatim}
where \var{looptools-path} is the path to your LoopTools source directory and
\var{defs} and \var{opts} are any additional variable definitions or options
you like to use when compiling LoopTools. After successfully re-compiling
LoopTools pass the option
\begin{Verbatim}[commandchars=\\\{\}]
--with-looptools=\var{looptools-path}
\end{Verbatim}
to the HEPMath \code{configure} script. Since LoopTools is written in Fortran
and the HEPMath interface in C the correct Fortran run-time libraries are needed
to link the interface. Usually \code{configure} will find them automatically,
but if you have several Fortran compilers installed on your system it might pick
the wrong one. (It needs the one you used to compile LoopTools, but it has no
way of detecting which one that was.) You can point it to the right compiler,
say, \code{gfortran} with the option
\begin{Verbatim}[commandchars=\\\{\}]
F77=gfortran
\end{Verbatim}
Alternatively, you can specify the options needed to compile and link C code
with LoopTools with the following options
\begin{Verbatim}[commandchars=\\\{\}]
LOOPTOOLS=\var{/path/to/libooptools.a}
LOOPTOOLS_CFLAGS=\var{compile-flags}
LOOPTOOLS_LIBS=\var{link-flags}
\end{Verbatim}
where \var{link-flags} should \emph{not} include the \code{-looptools} flag.

Note that HEPMath does not use the the Mathematica interface provided in the
LoopTools distribution, and you do not need to compile this interface in order
to access LoopTools from HEPMath. The reason for this duplication of interfaces
is that the one provided by LoopTools is based on MathLink. The code generation
method used in HEPMath can only handle calls to external functions if these
functions are interfaced via LibraryLink, and therefore HEPMath must implement
its own interface to LoopTools.

Once you have HEPMath installed correctly you can try and run the examples in
the \code{examples} sub-directory of the source directory. The provided examples
are:
\begin{description}
\item[\code{ee-mumu.m}] The computation of the $e^+e^-\to\mu^+\mu^-$ squared
  matrix element in QED. This example does not require FeynArts, LHAPDF or
  LoopTools.
\item[\code{hepcompile.m}] A short demonstration of the code generation
  mechanism, including the generation of a Python extension module.  This
  example requires the LoopTools interface (but you can easily modify it so that
  it doesn't).
\item[\code{H-gg-SM.m}] The computation of the top-loop-induced $H\to gg$ decay
  width in the SM at leading order. This example illustrates the use of the
  FeynArts interface, the computation of one-loop integrals with LoopTools,
  colour algebra and code generation. It requires FeynArts and the LoopTools
  interface.
\item[\code{gg-gH-SM.m}] The computation of the $gg\to gH$ process in the SM at
  leading order. This process is used as a performance benchmark in
  Sec.~\ref{sec:performance}. The script needs the FeynArts and LoopTools
  interfaces.
\end{description}
I recommend to run the files as scripts, i.e. with
\begin{Verbatim}[commandchars=\\\{\}]
math -script \var{example}.m
\end{Verbatim}
The source code in the examples is well commented and should get you started
quickly with HEPMath. For a more structured introduction just continue reading.

%%%%%%%%%%%%%%%%%%%%%%%%%%%%%%%%%%%%%%%%%%%%%%%%%%%%%%%%%%%%%%%%%%%%%%%%%%%%%%%%
\section{Building Blocks for Feynman Amplitudes}
\label{sec:buildingblocks}
%%%%%%%%%%%%%%%%%%%%%%%%%%%%%%%%%%%%%%%%%%%%%%%%%%%%%%%%%%%%%%%%%%%%%%%%%%%%%%%%

If HEPMath is properly installed you can open a Mathematica session and load
the package with
\begin{Verbatim}[commandchars=\\\{\}]
Needs["HEPMath`"]
\end{Verbatim}
The HEPMath package defines a number of symbols which represent typical
building blocks of Feynman amplitudes within Mathematica. This section
summarises these building blocks and describes their pre-defined behaviour.

%%%%%%%%%%%%%%%%%%%%%%%%%%%%%%%%%%%%%%%%%%%%%%%%%%%%%%%%%%%%%%%%%%%%%%%%%%%%%%%%
\subsection{Tensors and Dirac-Algebra}
%%%%%%%%%%%%%%%%%%%%%%%%%%%%%%%%%%%%%%%%%%%%%%%%%%%%%%%%%%%%%%%%%%%%%%%%%%%%%%%%

Dirac matrices are represented by the symbol \code{Ga}. Try evaluating
\begin{Verbatim}[commandchars=\\\{\}]
HEPTensorSignature[Ga]
\out {\lb}Lorentz, Dirac, Dirac{\rb}
\end{Verbatim}
(The text after the arrow `$\Rightarrow$' is the expression returned by
Mathematica.) This tells you that \code{Ga} is an object which carries three
indices: one Lorentz index and two Dirac indices. All objects which can carry
indices are called \emph{HEPTensors} in HEPMath, and the
\code{HEPTensorSignature} function gives you a list with the types of
indices you can attach. In particular, any variable not known to HEPMath is
a \emph{scalar}, i.e.\ has an empty tensor signature.
\begin{Verbatim}[commandchars=\\\{\}]
HEPTensorSignature[xyz]
\out {\lb}{\rb}
\end{Verbatim}
To attach symbolic indices to \code{Ga} simply apply it to the list of indices.
\begin{Verbatim}[commandchars=\\\{\}]
HEPTensorSignature[Ga[mu, alpha, beta]]
\out {\lb}{\rb}
HEPTensorSignature[Ga[mu]]
\out {\lb}Dirac, Dirac{\rb}
\end{Verbatim}
Attaching the full list of indices turns \code{Ga} into a scalar, i.e.\ into an
object with empty tensor signature. By attaching an incomplete list of indices
you reduce the rank of the tensor, removing elements from its tensor signature
from the left. The full set of pre-defined tensors provided by HEPMath is
summarised in Tab.~\ref{tab:tens-predef}.
\ctable[
  caption = Pre-defined tensors and tensor-valued functions.,
  width=\textwidth,
  label=tab:tens-predef
]{>{\ttfamily}l>{\ttfamily}p{0.36\textwidth}>{\raggedright}X}{
  \tnote{\var{p} must be an expression with tensor signature
    \code{\{Lorentz\}}.}
}{
  \FL
  \textrm{Symbol} & \textrm{Tensor signature} & Description \ML
  Ga & \{Lorentz, Dirac, Dirac\} & The Dirac matrix $\gamma^\mu$. \NN
  GI & \{Dirac, Dirac\} & The identity matrix $\id$. \NN
  G5 & \{Dirac, Dirac\} & The matrix $\gamma_5$. \NN
  PL & \{Dirac, Dirac\} & $(\id-\gamma_5)/2$. \NN
  PR & \{Dirac, Dirac\} & $(\id+\gamma_5)/2$. \NN
  Gs[\var{p}]\tmark & \{Dirac, Dirac\} &
    The contraction $\gamma^\mu p_\mu\equiv\pslash$. \NN
  GSig & \{Lorentz, Lorentz,\newline\mbox{\ }Dirac, Dirac\} &
    The matrix $\sigma^{\mu\nu} =
    \tfrac{i}{2}(\gamma^\mu\gamma^\nu-\gamma^\nu\gamma^\mu)$. \NN
  USp[\var{p}]\tmark & \{FermionHelicity, Dirac\} & The Dirac spinor $u(p)$. \NN
  VSp[\var{p}]\tmark & \{FermionHelicity, Dirac\} & The Dirac spinor $v(p)$. \NN
  Pol[\var{p}]\tmark & \{VectorPolarization,\newline\mbox{\ }Lorentz\} &
    The massless polarisation vector $\epsilon^\mu(p)$. \NN
  MPol[\var{p}]\tmark & \{MassiveVectorPolarization,\newline\mbox{\ }Lorentz\} &
    The massive polarisation vector $\epsilon^\mu(p)$. \NN
  Dim & \{\} & Symbol representing the dimension of Minkowski space. \NN
  Eta & \{Lorentz, Lorentz\} &
    The metric tensor $\eta^{\mu\nu}\equiv g^{\mu\nu}$ in \code{Dim} dimensions
    with ``mostly minus'' signature. \NN
  EtaHat & \{Lorentz, Lorentz\} &
    The $(\code{Dim}-4)$-dimensional part of the metric tensor. \NN
  Eps & \{Lorentz, Lorentz,\newline\mbox{\ }Lorentz, Lorentz\} &
    The Levi-Civita tensor $\epsilon^{\mu\nu\rho\sigma}$. \NN
  Den[p, m]\tmark & \{\} & The propagator denominator \mbox{$1/(p^2-m^2)$}. \NN
  ColorT & \{ColorAdjoint,\newline\mbox{\ }%
             ColorFundamental,\newline\mbox{\ }%
             ColorFundamental\} &
    The $SU(3)$ generator $T^a$. \NN
  ColorF & \{ColorAdjoint,\newline\mbox{\ }%
             ColorAdjoint,\newline\mbox{\ }%
             ColorAdjoint\} &
    The $SU(3)$ structure constants $f^{abc}$. \NN
  ColorDelta & \{ColorFundamental,\newline\mbox{\ }%
                 ColorFundamental\} &
    The identity matrix for colour indices. \NN
  GluonDelta & \{ColorAdjoint,\newline\mbox{\ }%
                 ColorAdjoint\} &
    The identity matrix for adjoint colour indices. \LL
}

You can form linear combinations of tensors with the same tensor signature
and use them in the same way as you would use a tensor symbol. For example
\begin{Verbatim}[commandchars=\\\{\}]
HEPTensorSignature[a GI + b G5]
\out {\lb}Dirac, Dirac{\rb}
HEPTensorSignature[(a GI + b G5)[al, bt]]
\out {\lb}{\rb}
\end{Verbatim}
(See Tab.~\ref{tab:tens-predef} for an explanation of \code{GI} and \code{G5}.)
If you try to add two tensors with different tensor signature or multiply two
tensors HEPMath will complain.
\begin{Verbatim}[commandchars=\\\{\}]
HEPTensorSignature[a Ga + b G5]
\out HEPTensorSignature::conflict:
\noout{}    Incompatible tensors in tensor expression b G5 + a Ga.
\out $Aborted
HEPTensorSignature[GI G5]
\out HEPTensorSignature::conflict:
\noout{}    Incompatible tensors in tensor expression GI G5.
\out $Aborted
\end{Verbatim}

Implicit summation over repeated indices is understood and fully supported by
HEPMath. You can get a summary of the free and contracted indices in an
expression with the \code{Indices} function.
\begin{Verbatim}[commandchars=\\\{\}]
Indices[Ga[mu, al, bt] Ga[mu, bt, ga]]
\out {\lb}{\lb}al -> Dirac, ga -> Dirac{\rb}, {\lb}bt -> Dirac, mu -> Lorentz{\rb}{\rb}
\end{Verbatim}
The first list holds the free indices of the expression and the second list
the contracted indices. Note how the types of the indices are inferred from
their position in the index list. You can also use the \code{Indices} function
to find out if an expression is well-formed:
\begin{Verbatim}[commandchars=\\\{\}]
Indices[Ga[mu, al, al] Ga[mu, al, al]]
\out Indices::mult: Index repeated more than twice in product
\noout{}    Ga[mu, al, al] Ga[nu, al, al].
\out \code{\$}Aborted
\end{Verbatim}

HEPMath provides a few index-free notations for contractions between
tensors and functions for switching between indexed and index-free notation.
Try evaluating the following expressions:
\begin{Verbatim}[commandchars=\\\{\}]
DiracContract[Ga[mu, al, bt] Ga[nu, bt, ga]]
\out ( Ga[mu] . Ga[nu] )[al, ga]
DiracContract[Ga[mu, al, bt] Ga[nu, bt, al]]
\out DiracTr[Ga[mu] . Ga[nu]]
\end{Verbatim}
As you see, Mathematica's dot operator (\code{Dot}) can be used to contract
Dirac matrices. More generally, it will contract tensors with at most two
indices of the same type in the obvious way. The expression \code{Ga[mu].Ga[nu]}
has tensor signature \code{\{Dirac, Dirac\}} and can therefore be indexed with
two Dirac indices. (Feel free to verify this with \code{HEPTensorSignature}.)
\code{DiracTr} represents the trace over Dirac indices and can be applied to any
object with tensor signature \code{\{Dirac, Dirac\}}. If you prefer to have all
indices written out explicitly you can undo the effect of \code{DiracContract}
with
\begin{Verbatim}[commandchars=\\\{\}]
ExplicitIndices[(Ga[mu] . Ga[nu])[al, ga]]
\out Ga[mu, al, $3] Ga[nu, $3, ga]
\end{Verbatim}
Note how HEPMath generates unique symbols for the dummy indices.

The Dirac trace can be calculated with the \code{CalcDiracTraces} function.
\begin{Verbatim}[commandchars=\\\{\}]
CalcDiracTraces[DiracTr[Ga[mu] . Ga[nu]]]
\out 4 Eta[mu, nu]
\end{Verbatim}
The symbol \code{Eta} represents the metric tensor $\eta^{\mu\nu}\equiv
g^{\mu\nu}$ in $D\equiv\code{Dim}$ dimensions with the ``mostly minus''
signature. The \code{CalcDiracTraces} function computes any occurrence of
\code{DiracTr[...]} in its argument and substitutes the result. Occurrences of
$\gamma_5$ matrices (represented by \code{G5}, see Tab.~\ref{tab:tens-predef})
are handled the same way as in TRACER, i.e. by letting the
$(D-4)$-dimensional part of $\gamma^\mu$ \emph{commute} with $\gamma_5$.
The option \code{AntiCommutingG5} lets you use a completely anti-commuting
$\gamma_5$ instead:
\begin{Verbatim}[commandchars=\\\{\}]
CalcDiracTraces[DiracTr[Ga[mu].G5.Ga[nu].G5]]
\out -4 Eta[mu, nu] + 8 EtaHat[mu, nu]
CalcDiracTraces[DiracTr[Ga[mu].G5.Ga[nu].G5], AntiCommutingG5->True]
\out -4 Eta[mu, nu]
\end{Verbatim}

The symbol \code{EtaHat} represents the $(D-4)$-dimensional part of the
metric tensor. You can check the dimension of \code{Eta} and \code{EtaHat}
with the \code{LorentzContract} function:
\begin{Verbatim}[commandchars=\\\{\}]
LorentzContract[Eta[mu,mu]]
\out Dim
LorentzContract[EtaHat[mu,mu]]
\out -4 + Dim
\end{Verbatim}
The Levi-Civita tensor is represented by the symbol \code{Eps}. Its indices live
in 4 dimensions.
\begin{Verbatim}[commandchars=\\\{\}]
LorentzContract[Eps[mu, nu, ro, sg] EtaHat[sg, lm]]
\out 0
\end{Verbatim}
You can use the \code{SortEps} function to bring the arguments of \code{Eps} in
canonical order (as defined by Mathematica's \code{OrderedQ} function)
\begin{Verbatim}[commandchars=\\\{\}]
SortEps[a Eps[nu, mu, ro, sg] + b Eps[mu, nu, ro, nu]]
\out -a Eps[mu, nu, ro, sg]
\end{Verbatim}

%%%%%%%%%%%%%%%%%%%%%%%%%%%%%%%%%%%%%%%%%%%%%%%%%%%%%%%%%%%%%%%%%%%%%%%%%%%%%%%%
\subsection{Lorentz Vectors and Dirac Spinors}
\label{sec:lorentz}
%%%%%%%%%%%%%%%%%%%%%%%%%%%%%%%%%%%%%%%%%%%%%%%%%%%%%%%%%%%%%%%%%%%%%%%%%%%%%%%%

Playing with the pre-defined tensors of HEPMath is nice, but for an actual
calculation you also need to be able to define your own tensors. At the very
least you will need symbols that represent Lorentz vectors. To tell HEPMath
that a certain symbol represents a Lorentz vector you have to declare it.
\begin{Verbatim}[commandchars=\\\{\}]
DeclareLorentzVectors[p]
DeclareLorentzVectors[p, k, q]
DeclareLorentzVectors[{\lb}p, k, q{\rb}]
\end{Verbatim}
These commands tell HEPMath that the symbols \code{p}, \code{k}, and \code{q}
represent \emph{real, $D$-dimensional} Lorentz vectors. The arguments of
\code{DeclareLorentzVectors} can be arbitrary \emph{patterns}. Any expression
matching one of these patterns will be recognised as a Lorentz vector. For
example
\begin{Verbatim}[commandchars=\\\{\}]
DeclareLorentzVectors[_p]
\end{Verbatim}
makes any expression of the form \code{p[\var{something}]} a Lorentz vector.

Complex and/or four-dimensional Lorentz vectors can be declared with the
functions
\begin{Verbatim}[commandchars=\\\{\}]
Declare4DLorentzVectors
DeclareComplexLorentzVectors
DeclareComplex4DLorentzVectors
\end{Verbatim}
Similar to \code{DiracContract} the \code{LorentzContract} function can be used
to eliminate unnecessary occurrences of \code{Eta} and to write contractions
between Lorentz vectors in index-free form.
\begin{Verbatim}[commandchars=\\\{\}]
LorentzContract[Eta[mu, nu] p[nu]]
\out p[mu]
LorentzContract[k[mu] p[mu]]
\out SDot[k, p]
\end{Verbatim}
The last expression is equivalent to \code{k.p} but holds the additional
information that the product is symmetric. (The \code{SDot} symbol has the
attribute \code{Orderless}.) Although mostly used for products of Lorentz
vectors, \code{SDot} is more general and can represent the full contraction of
any two tensors with the same tensor signature.
\begin{Verbatim}[commandchars=\\\{\}]
ExplicitIndices[SDot[Eta, EtaHat]]
\out Eta[$3, $4] EtaHat[$3, $4]
\end{Verbatim}
The full contraction of a tensor with itself is represented by the \code{Sqr}
function:
\begin{Verbatim}[commandchars=\\\{\}]
SDot[p, p]
\out Sqr[p]
ExplicitIndices[Sqr[p]]
\out p[$3]^2
\end{Verbatim}
Note that HEPMath is aware that the index \code{\$3} in \code{p[\$3]\^{}2}
is contracted.
\begin{Verbatim}[commandchars=\\\{\}]
Indices[p[\code{\$}3]^2]
\out {\lb}{\lb}{\rb}, {\lb}\code{\$}3 -> Lorentz{\rb}{\rb}
\end{Verbatim}

The distinguishing feature of four-dimensional Lorentz vectors is that their
contractions with \code{EtaHat} vanish:
\begin{Verbatim}[commandchars=\\\{\}]
Declare4DLorentzVectors[p];
LorentzContract[EtaHat[mu, nu] p[nu]]
\out 0
\end{Verbatim}

HEPMath also defines a few ``tensor-valued functions'', i.e. symbols which
become a tensor only when they are applied to something else. For example,
Dirac spinors $u(p)$ and $v(p)$ associated with a four-momentum $p$
can be represented by \code{USp[p]} and \code{VSp[p]}, respectively.
\begin{Verbatim}[commandchars=\\\{\}]
DeclareLorentzVectors[p];
HEPTensorSignature[USp[p]]
\out {\lb}FermionHelicity, Dirac{\rb}
HEPTensorSignature[VSp[p]]
\out {\lb}FermionHelicity, Dirac{\rb}
\end{Verbatim}
Note that the helicities of the spinors are represented by the first index whose
type is \code{FermionHelicity}. Polarisation vectors of massless or massive
vector bosons can be represented by \code{Pol} and \code{MPol} in an analogous
way.
\begin{Verbatim}[commandchars=\\\{\}]
DeclareLorentzVectors[p];
HEPTensorSignature[Pol[p]]
\out {\lb}VectorPolarization, Lorentz{\rb}
HEPTensorSignature[MPol[p]]
\out {\lb}MassiveVectorPolarization, Lorentz{\rb}
\end{Verbatim}

Contractions of Lorentz vectors with the Levi-Civita tensor are represented by
\code{EpsDot}. It takes up to four Lorentz vectors as arguments and its tensor
signature depends on the number of supplied arguments.
\begin{Verbatim}[commandchars=\\\{\}]
DeclareLorentzVectors[k,l,p,q];
LorentzContract[Eps[mu, nu, ro, sg] k[mu] l[nu] p[ro] q[sg]]
\out EpsDot[k, l, p, q]
LorentzContract[Eps[mu, nu, ro, sg] k[mu] l[nu] p[ro]]
\out EpsDot[k, l, p][sg]
LorentzContract[Eps[mu, nu, ro, sg] k[mu] l[nu]]
\out EpsDot[k, l][ro, sg]
LorentzContract[Eps[mu, nu, ro, sg] k[mu]]
\out EpsDot[k][mu, ro, sg]
\end{Verbatim}
As for \code{Eps}, you can use the \code{SortEps} function to bring \code{EpsDot}
expressions in canonical order.
\begin{Verbatim}[commandchars=\\\{\}]
SortEps[a EpsDot[p, q][nu, mu] + b EpsDot[q, p][mu, nu] + 
        c EpsDot[p, p][mu, nu] + d EpsDot[p, q][mu, mu]]
\out -a*EpsDot[p, q][mu, nu] - b*EpsDot[p, q][mu, nu]
SortEps[EpsDot[l, k, p, q]]
\out -EpsDot[k, l, p, q]
\end{Verbatim}

%%%%%%%%%%%%%%%%%%%%%%%%%%%%%%%%%%%%%%%%%%%%%%%%%%%%%%%%%%%%%%%%%%%%%%%%%%%%%%%%
\subsection{Conjugation}
%%%%%%%%%%%%%%%%%%%%%%%%%%%%%%%%%%%%%%%%%%%%%%%%%%%%%%%%%%%%%%%%%%%%%%%%%%%%%%%%

Complex conjugation is represented in HEPMath by the function \code{CC}.
Unlike Mathematica's \code{Conjugate} function it will automatically distribute
itself over sums and products and it will automatically disappear when acting
on a symbol that has been declared as real. To declare a symbol as real you can
use the \code{DeclareReal} function.
\begin{Verbatim}[commandchars=\\\{\}]
DeclareLorentzVectors[p];
DeclareComplexLorentzVectors[e];
DeclareReal[r];
CC[c p[mu] + r e[mu]]
\out CC[c] p[mu] + r CC[e][mu]
\end{Verbatim}
Note that in the second term \code{CC} is applied to the vector \code{e},
not the indexed vector \code{e[mu]}.

The `bar' notation for Dirac spinors and matrices is represented by the function
\code{Bar}. Like \code{CC} it distributes itself over sums and products.
The relationship between complex conjugation and bar conjugation is fully
implemented:
\begin{Verbatim}[commandchars=\\\{\}]
Bar[r USp[p][s] + c VSp[p][s]]
\out r Bar[USp[p]][s] + CC[c] Bar[VSp[p]][s]
CC[Bar[USp[p]][s].PL.Gs[p].PR.Gs[e].G5.VSp[p][s]]
\out Bar[VSp[p]][s].(-G5).Gs[CC[e]].PL.Gs[p].PR.USp[p][s]
\end{Verbatim}
Note that I attached an explicit spin index \code{s} to the Dirac spinors above,
but suppressed the Dirac indices.

%%%%%%%%%%%%%%%%%%%%%%%%%%%%%%%%%%%%%%%%%%%%%%%%%%%%%%%%%%%%%%%%%%%%%%%%%%%%%%%%
\subsection{Colour Algebra}
%%%%%%%%%%%%%%%%%%%%%%%%%%%%%%%%%%%%%%%%%%%%%%%%%%%%%%%%%%%%%%%%%%%%%%%%%%%%%%%%

$SU(3)$ generators $T^a$ and structure constants $f^{abc}$ are represented in
HEPMath by the \code{ColorT} and \code{ColorF} tensors, respectively. The index
types for the fundamental and adjoint $SU(3)$ representation are called
\code{ColorFundamental} and \code{ColorAdjoint}. The identity matrices in these
two index spaces are called \code{ColorDelta} and \code{GluonDelta},
respectively. There is no equivalent for \code{DiracContract} or
\code{LorentzContract}, but you can remove unnecessary appearances of identity
matrices (in any index space) with the \code{HEPContract} function.
\begin{Verbatim}[commandchars=\\\{\}]
HEPContract[ColorDelta[i, j] ColorT[a, j, k]]
\out T[a, i, k]
HEPContract[GluonDelta[a, b] ColorF[b, c, d]]
\out ColorF[a, c, d]
\end{Verbatim}

Traces over fundamental colour indices are represented by the \code{ColorTr}
function, and the hermicity of the $SU(3)$ generators is implemented
consistently.
\begin{Verbatim}[commandchars=\\\{\}]
CC[ColorT[a, i, j]]
\out T[a, j, i]
CC[ColorTr[ColorT[a].ColorT[b].ColorT[c]]]
\out ColorTr[ColorT[c] . ColorT[b] . ColorT[a]]
\end{Verbatim}

%%%%%%%%%%%%%%%%%%%%%%%%%%%%%%%%%%%%%%%%%%%%%%%%%%%%%%%%%%%%%%%%%%%%%%%%%%%%%%%%
\section{Algebraic Manipulations}
\label{sec:algebra}
%%%%%%%%%%%%%%%%%%%%%%%%%%%%%%%%%%%%%%%%%%%%%%%%%%%%%%%%%%%%%%%%%%%%%%%%%%%%%%%%

In the last section I described the building blocks which HEPMath provides in
order to build expressions representing Feynman amplitudes. Now it is time to
learn how to put them together and manipulate them. To do this you first need to
know which kind of constructs are ``understood'' by the HEPMath system.
HEPMath follows a design principle which I call \emph{substitution invariance}
and which can be summarised by the following statement: \emph{substituting valid
  expressions by other valid expressions with the same tensor signature in valid
  expressions must yield valid expressions.} What this means is simple. Take the
expression \code{p[mu]} which represents a Lorentz vector \code{p} with an index
\code{mu} attached to it. When we replace \code{p} (a valid expression with
tensor signature \code{\{Lorentz\}}) by the sum \code{k+q} of two other vectors
(also a valid expression with tensor signature \code{\{Lorentz\}}) we get
\code{(k+q)[mu]}. So, by substitution invariance \code{(k+q)[mu]} must be a
valid expression if \code{p[mu]} is valid. The importance of substitution
invariance is evident when you imagine some complicated expression in which you
want to substitute \code{p} by \code{k+q}. The vector \code{p} can appear in
various forms: \code{p[mu]}, \code{Sqr[p]}, \code{Gs[p]}, \code{USp[p][s]}
etc. Writing a separate replacement rule for each of these cases is very
error-prone and likely to make your code extremely hard to read. Therefore,
HEPMath must be able to handle whatever expressions arise from the single
substitution \code{p -> k+q}.

The main consequence of substitution invariance is that the following
``funny'' expressions must be valid:
\begin{itemize}
\item composite heads: \code{(a k + b q)[mu]},
\item multiple index lists: \code{Ga[mu][al, bt]}.
\end{itemize}
Standard Mathematica functions for algebraic manipulations such as \code{Expand}
or \code{Collect} obviously don't know what to do with these expressions.
HEPMath therefore complements Mathematica's library of symbolic manipulation
functions with functions that can handle tensor expressions, including the
``funny'' ones above. In addition it implements several ``HEP-specific''
functions which are frequently needed in High Energy Physics computations. This
section describes both types of functions, starting with the basic ones. Some of
the functions have already been mentioned in Sec.~\ref{sec:buildingblocks}, but I
list them here again for completeness.

%%%%%%%%%%%%%%%%%%%%%%%%%%%%%%%%%%%%%%%%%%%%%%%%%%%%%%%%%%%%%%%%%%%%%%%%%%%%%%%%
\subsection{Expanding and Collecting}
%%%%%%%%%%%%%%%%%%%%%%%%%%%%%%%%%%%%%%%%%%%%%%%%%%%%%%%%%%%%%%%%%%%%%%%%%%%%%%%%

\begin{description}

\item[\code{HEPExpand}]
mimics Mathematica's \code{Expand} function.  Like \code{Expand} it takes a
pattern as an optional second argument. All sub-expressions which are free of
this pattern are not expanded. However, unlike \code{Expand}, \code{HEPExpand}'s
default behaviour is to leave sub-expressions alone which do not contain any
HEPTensors.
\begin{Verbatim}[commandchars=\\\{\}]
DeclareLorentzVectors[p, q];
HEPExpand[((a + b) (p + q))[mu]]
\out (a + b) p[mu] + (a + b) q[mu]
HEPExpand[(a (p + q))[mu], p]
\out a p[mu] + (a q)[mu]
\end{Verbatim}
Note how in the second case the factor \code{a} has not been pulled out of the
head of the expression \code{(a q)[mu]} since this expression does not contain
\code{p}. Needless to say, \code{HEPExpand} operates correctly on traces and
index-free contractions such as \code{Gs}, \code{Sqr}, \code{SDot} and
\code{Dot}.
\begin{Verbatim}[commandchars=\\\{\}]
HEPExpand[Bar[USp[p]][s].(Gs[p + q] + m GI).VSp[q][s]]
\out m Bar[USp[p]][s] . GI . VSp[q][s] + 
\noout{}    Bar[USp[p]][s] . Gs[p] . VSp[q][s] +
\noout{}    Bar[USp[p]][s] . Gs[q] . VSp[q][s]
HEPExpand[DiracTr[Ga[mu].(a GI + b G5).Ga[nu]]]
\out b DiracTr[Ga[mu] . G5 . Ga[nu]] + a DiracTr[Ga[mu] . GI . Ga[nu]]
HEPExpand[Sqr[p + q]]
\out 2 SDot[p, q] + Sqr[p] + Sqr[q]
\end{Verbatim}

\item[\code{PullScalars}]
Sometimes you just want to pull scalar factors from the heads of indexed
expressions or linear functions like \code{SDot} without expanding everything.
HEPMath gives you the function \code{PullScalars} to do that. 
\begin{Verbatim}[commandchars=\\\{\}]
PullScalars[(a (p + q))[mu]]
\out a (p + q)[mu]
PullScalars[DiracTr[Gs[a (p + q)].Gs[Sqr[p] q]], _Sqr]
\out DiracTr[Gs[a (p + q)] . Gs[q]] Sqr[p]
\end{Verbatim}
The optional second argument can be any Mathematica pattern. Only scalars
matching the pattern are pulled out.

\item[\code{SortEps}]
was already mentioned in Sec.~\ref{sec:lorentz}. It uses anti-symmetry of the
Levi-Civita tensor to put the arguments of \code{Eps} and \code{EpsDot}
expressions in canonical order.

\item[\code{HEPCollect}]
Factorising expressions in HEPMath is complicated by the fact that tensor
expressions can be put together with more operators than addition and
multiplication. Just remember the different forms in which a Lorentz vector can
appear: \code{p[mu]}, \code{Sqr[p]}, \code{SDot[p, q]} etc. The
\code{HEPCollect} function implements a behaviour similar to FORM's
\code{bracket} statement. \code{HEPCollect[expr, p]} goes through each term
in \code{expr} and finds the factors which contain \code{p}. It then introduces
one level of brackets, combining those terms which have the same set of
\code{p}-dependent factors. The second argument \code{p} can be a Mathematica
pattern, in which case all factors which are not free of this pattern are
pulled out.
\begin{Verbatim}[commandchars=\\\{\}]
expr = SDot[p, q] a1 + SDot[p, q] a2 +
       Sqr[p] b1 + Sqr[p] b2 +
       SDot[p, q] Sqr[p] c1 + SDot[p, q] Sqr[p] c2;
HEPCollect[expr, p]
\out (a1 + a2) SDot[p, q] + (b1 + b2) Sqr[p] + 
\noout (c1 + c2) SDot[p, q] Sqr[p]
HEPCollect[expr, q]
\out b1 Sqr[p] + b2 Sqr[p] + 
\noout SDot[p, q] (a1 + a2 + c1 Sqr[p] + c2 Sqr[p])
\end{Verbatim}
By default \code{HEPCollect} automatically renames contracted indices to
minimise the number of generated terms:
\begin{Verbatim}[commandchars=\\\{\}]
HEPCollect[p[mu] k[mu] + p[nu] q[nu], p]
\out p[mu] (k[mu] + q[mu])
\end{Verbatim}
You can disable this feature (e.g. for performance reasons) by setting the option
\code{CombineTensorStructures} to \code{False}.

\item[\code{HEPCollectListed}]
A common use-case for functions like \code{HEPCollect} is the task of isolating
different colour or Lorentz structures in an expression and extracting the
coefficients of these structures. Typically you then want to perform some
operations only on the structures or only on the coefficients. The function
\code{HEPCollectListed} is made for this purpose since it returns the structures
and coefficients as a pair of lists.
\begin{Verbatim}[commandchars=\\\{\}]
expr = a Eta[mu, nu] + b Sqr[p] Eta[mu, nu] + c p[mu] p[nu];
terms = HEPCollectListed[expr, mu|nu]
\out {\lb}{\lb}Eta[mu, nu], p[mu] p[nu]{\rb}, {\lb}a + b Sqr[p], c{\rb}{\rb}
\end{Verbatim}
In fact, the \code{HEPCollect} function is simply implemented as
\begin{Verbatim}[commandchars=\\\{\}]
HEPCollect[expr_, patt_] := Dot @@ HEPCollectListed[expr, patt]
\end{Verbatim}

\end{description}

Pairs of lists as returned by \code{HEPCollectListed} can be processed further
with the \code{HEPCoefficient}, \code{HEPSelectTerms}, and \code{HEPGroupTerms}
functions.

\begin{description}
\item[\code{HEPCoefficient}] can be used to extract the coefficient of a
specific term. Continuing the example above, we can extract the coefficient
of the \code{Eta[mu, nu]} term with
\begin{Verbatim}[commandchars=\\\{\}]
HEPCoefficient[terms, Eta[mu, nu]]
\out a + b Sqr[p]
\end{Verbatim}
You can use patterns to identify the desired term:
\begin{Verbatim}[commandchars=\\\{\}]
HEPCoefficient[terms, _Eta]
\out a + b Sqr[p]
\end{Verbatim}
However, you will receive an error message if the pattern matches more than one
term.

\item[\code{HEPSelectTerms}] Similar to Mathematica's \code{Select} function you
can use \code{HEPSelectTerms} to filter out those terms for which a given
function returns \code{True}:
\begin{Verbatim}[commandchars=\\\{\}]
HEPSelectTerms[terms, Head[#] === Eta &]
\out {\lb}{\lb}Eta[mu, nu]{\rb}, {\lb}a + b Sqr[p]{\rb}{\rb}
\end{Verbatim}
The result is of the same form as the results of \code{HEPCollectListed}.

\item[\code{HEPGroupTerms}] If you want to split the terms into two sets
according to some criterium you can do that with the \code{HEPGroupTerms}
function. It returns a list of two expressions of the form returned by
\code{HEPCollectListed}.
\begin{Verbatim}[commandchars=\\\{\}]
HEPGroupTerms[terms, Head[#] === Eta &]
\out {\lb}{\lb}{\lb}Eta[mu, nu]{\rb}, {\lb}a + b Sqr[p]{\rb}{\rb},
\noout{}  {\lb}{\lb}p[mu] p[nu]{\rb}, {\lb}c{\rb}{\rb}{\rb}
\end{Verbatim}

\end{description}

In addition to the above-mentioned ``generic'' tensor manipulation functions
HEPMath also provides a few functions which are specific to Lorentz vectors
and Dirac matrices.

\begin{description}

\item[\code{DiracSubstitute}]
replaces occurrences of \code{PL}, \code{PR} and \code{GSig[mu, nu]} by their
definitions in terms of \code{GI}, \code{G5} and \code{Ga[mu]}.

\item[\code{DiracExpand}] can be used to expand composite factors in
  \code{DiracTr} expressions:
\begin{Verbatim}[commandchars=\\\{\}]
DiracExpand[DiracTr[(Gs[p] + m GI).PL]]
\out (-1/2) m DiracTr[G5] + (1/2) m DiracTr[GI] - 
\noout (1/2) DiracTr[Gs[p].G5] + (1/2) DiracTr[Gs[p]]
\end{Verbatim}
It is defined as
\begin{Verbatim}[commandchars=\\\{\}]
DiracExpand[expr_] := 
    DiracContract[HEPExpand[DiracSubstitute[expr], Gs|Ga|GI|G5]]
\end{Verbatim}

\item[\code{PullVectors}]
is similar to \code{PullScalars} but extracts Lorentz vectors from index-free
contractions like \code{Gs[p]} or \code{SDot[p, q]}. Such an operation is
often needed in reduction algorithms for loop integrals.
\begin{Verbatim}[commandchars=\\\{\}]
PullVectors[DiracTr[(Gs[p] + m GI).(SDot[p, q] G5)], p]
\out DiracTr[(GI m) . G5] p[$3] q[$3] + 
\noout DiracTr[Ga[$4] . G5] p[$4] p[$5] q[$5]
\end{Verbatim}
The optional second argument can be any Mathematica pattern. Only vectors
matching the pattern are pulled out. \code{PullVectors[expr, p]} expands
\code{expr} in \code{p}, i.e.\ it calls \code{HEPExpand[expr, p]}.

\end{description}

%%%%%%%%%%%%%%%%%%%%%%%%%%%%%%%%%%%%%%%%%%%%%%%%%%%%%%%%%%%%%%%%%%%%%%%%%%%%%%%%
\subsection{Index Manipulations and Substitutions}
%%%%%%%%%%%%%%%%%%%%%%%%%%%%%%%%%%%%%%%%%%%%%%%%%%%%%%%%%%%%%%%%%%%%%%%%%%%%%%%%

\begin{description}

\item[\code{JoinIndexLists}] can be used to join multiple index lists.
\begin{Verbatim}[commandchars=\\\{\}]
JoinIndexLists[USp[p][s][al] + G5[al][bt] VSp[q][s][bt]]
\out USp[p][s, al] + G5[al, bt] VSp[q][s, bt]
\end{Verbatim}

\item[\code{HEPContract}]
Identity tensors such as \code{Eta} (for Lorentz indices), \code{GI} (for Dirac
indices) or \code{ColorDelta} (for colour indices) can be eliminated with the
\code{HEPContract} function. Full contractions of identity tensors with
themselves are replaced by the correct dimension.
\begin{Verbatim}[commandchars=\\\{\}]
HEPContract[Eta[mu][nu] p[nu]]
\out p[mu]
HEPContract[Ga[mu].GI.Ga[nu]]
\out Ga[mu] . Ga[nu]
HEPContract[a Eta[mu, mu] + b GI[al, al] + c ColorDelta[i, i]]
\out 4 b + 3 c + a Dim
\end{Verbatim}
Note that \code{HEPContract} invokes \code{JoinIndexLists}.

\item[\code{LorentzContract}, \code{DiracContract}]
were already mentioned in Sec.~\ref{sec:buildingblocks}. They invoke
\code{HEPContract} and then perform additional replacements specific to the
built-in tensors.

\item[\code{ExplicitIndices}]
replaces index-free notations like \code{SDot[p, q]} with explicit index
contractions. It only works on expressions without suppressed indices.

\item[\code{ExplicitSlash}]
writes the contraction \code{Gs[p]} with explicit indices. This function also
works on expressions with suppressed indices:
\begin{Verbatim}[commandchars=\\\{\}]
ExplicitSlash[Gs[p]]
\out Ga[$3] p[$3]
\end{Verbatim}

\item[\code{HEPMultiply}]
When dealing with tensor expressions name clashes between contracted indices are
a common problem. Consider the expressions \code{p[mu] k[mu]} and \code{q[mu]
  k[mu]}. By themselves they are absolutely fine, but if you want to multiply
them you have to re-name the contracted index \code{mu} in one of them.  For
complicated expressions the \code{Indices} function mentioned in
Sec.~\ref{sec:buildingblocks} can help you find all contracted
indices. Alternatively, you can let HEPMath do the renaming automatically by
using the \code{HEPMultiply} function. It replaces contracted indices in all its
arguments by unique symbols and then multiplies them.
\begin{Verbatim}[commandchars=\\\{\}]
HEPMultiply[p[mu] k[mu], q[mu] k[mu]]
\out k[$3] k[$4] p[$3] q[$4]
\end{Verbatim}
A common application is squaring a Feynman amplitude.
\begin{Verbatim}[commandchars=\\\{\}]
squaredme = HEPMultiply[amp, CC[amp]]
\end{Verbatim}

\item[\code{HEPReplaceAll}]
Another case where name clashes between contracted indices can occur are 
substitutions. Take the expression \code{a p[mu] k[mu]}. Replacing \code{a}
with \code{q[mu] k[mu]} using Mathematica's \code{ReplaceAll} function
(also known as the \code{/.} operator) will give you an invalid expression.
HEPMath can take care of this if you use the \code{HEPReplaceAll} function
instead.
\begin{Verbatim}[commandchars=\\\{\}]
HEPReplaceAll[p[mu] k[mu] a, a -> q[mu] k[mu]]
\out k[mu] k[$3] p[mu] q[$3]
\end{Verbatim}

\item[\code{ReplaceIndices}]
If you simply want to replace all contracted indices in an expression with
unique symbols you can use the \code{ReplaceIndices} function.
\begin{Verbatim}[commandchars=\\\{\}]
ReplaceIndices[p[mu] k[mu]]
\out p[$3] k[$3]
\end{Verbatim}

\item[\code{HEPSetZero}]
Setting a tensor variable to zero, either globally or by replacement rules, can
lead ill-formed expressions. For example, the replacement \code{p -> 0}
transforms the expression \code{p[mu] k[mu]} into the somewhat undesirable
result \code{0[mu] k[mu]}. Unfortunately there is no way to tell Mathematica to
automatically simplify such expressions to zero. \code{HEPSetZero[expr, p]} sets
the tensor \code{p} to zero inside the expression \code{expr} without generating
ill-formed expressions and returns the result:
\begin{Verbatim}[commandchars=\\\{\}]
HEPSetZero[Sqr[k] + p[mu] k[mu], p]
\out Sqr[k]
\end{Verbatim}
The second argument can also be a pattern, in which case all tensors matching
that pattern are set to zero.

\end{description}

%%%%%%%%%%%%%%%%%%%%%%%%%%%%%%%%%%%%%%%%%%%%%%%%%%%%%%%%%%%%%%%%%%%%%%%%%%%%%%%%
\subsection{Tensor Derivatives}
%%%%%%%%%%%%%%%%%%%%%%%%%%%%%%%%%%%%%%%%%%%%%%%%%%%%%%%%%%%%%%%%%%%%%%%%%%%%%%%%

Derivatives with respect to tensor variables can be computed with the
\code{HEPD} function. It mimicks the behaviour of Mathematica's \code{D}
function.
\begin{Verbatim}[commandchars=\\\{\}]
HEPD[\var{expr}, \var{t}[\var{i1},\ldots{},\var{in}]]
\end{Verbatim}
returns the derivative of expression \var{expr} with respect to the
tensor variable \var{t} with indices \var{i1},\ldots,\var{in}. This derivative
would usually be denoted
\begin{equation*}
  \frac{\partial\var{expr}}{\partial t_{i_1\cdots i_n}}
  \eqsep.
\end{equation*}
The index list must be complete, i.e.\ the expression
\code{\var{t}[\var{i1},\ldots{},\var{in}]} must have tensor signature
\code{\{\}}. The indices \var{i1},\ldots,\var{in} of the derivative must be
different from any contracted indices in \var{expr}. They can, however, be the
same as an external index of \var{expr}, in which case the derivative index is
contracted with the external index of \var{expr}. They can also be contracted
with each other. Higher derivatives can be computed by simply adding arguments
to \code{HEPD}
\begin{Verbatim}[commandchars=\\\{\}]
HEPD[\var{expr}, t[i1,\ldots{},in], r[j1,\ldots{},jm], \ldots{}]
\end{Verbatim}

The \code{HEPD} function can handle sums, products, powers, and index-free
contractions (via \code{SDot} and \code{Sqr}) of arbitrary tensor expressions.
More complicated constructs such as \code{Gs} or \code{DiracTr} are currently
not supported. (But you can always use \code{ExplicitIndices} to obtain an
expression that \code{HEPD} can handle.) When \code{HEPD} encounters an
expression it cannot differentiate it returns a \code{HEPDerivative} expression:
\begin{Verbatim}[commandchars=\\\{\}]
HEPD[Gs[p][al, bt], p[mu]]
\out HEPDerivative[Gs[p][al, bt], p[mu]]
\end{Verbatim}
Thus, tensor derivatives of user-defined functions can be set by specifying
rules for \code{HEPDerivative}.

%%%%%%%%%%%%%%%%%%%%%%%%%%%%%%%%%%%%%%%%%%%%%%%%%%%%%%%%%%%%%%%%%%%%%%%%%%%%%%%%
\subsection{Simplification of Colour Structures}
%%%%%%%%%%%%%%%%%%%%%%%%%%%%%%%%%%%%%%%%%%%%%%%%%%%%%%%%%%%%%%%%%%%%%%%%%%%%%%%%

To simplify colour structures and compute colour factors you can use the
\code{ColorReduce} function. It replaces occurrences of \code{ColorF}
with appropriate traces over the generators $T^a$ and then applies the
identity
\begin{equation*}
  T^a_{ij} T^b_{kl} = \tfrac12\delta_{il} \delta_{kj}
    - \tfrac16\delta_{ij} \delta_{kl}
  \eqsep.
\end{equation*}
Note that it only operates on expressions with no suppressed colour or gluon
indices.
\begin{Verbatim}[commandchars=\\\{\}]
ColorReduce[ColorF[a, b, c]]
\out (-2 I) ColorTr[ColorT[a] . ColorT[b] . ColorT[c]] +
\noout{}    (2 I) ColorTr[ColorT[a] . ColorT[c] . ColorT[b]]
ColorReduce[(ColorT[a].ColorT[a])[i, j]]
\out (4/3) ColorDelta[i, j]
ColorReduce[ColorT[a].ColorT[a]]
\out ColorReduce::nosup: ColorReduce only works on expressions 
\noout{}    without suppressed ColorFundamental and ColorAdjoint indices.
\out ColorT[a] . ColorT[a]
ColorReduce[ColorF[a, b, c] ColorF[a, b, c]]
\out 24
\end{Verbatim}

%%%%%%%%%%%%%%%%%%%%%%%%%%%%%%%%%%%%%%%%%%%%%%%%%%%%%%%%%%%%%%%%%%%%%%%%%%%%%%%%
\subsection{Polarisation Sums}
%%%%%%%%%%%%%%%%%%%%%%%%%%%%%%%%%%%%%%%%%%%%%%%%%%%%%%%%%%%%%%%%%%%%%%%%%%%%%%%%

The summation of a squared amplitude over the polarisations of external fermions
or vector bosons can be done analytically with the \code{FermionPolarizationSums}
and \code{VectorPolarizationSums} functions, respectively. The
\code{FermionPolarizationSums} function uses the formulae
\begin{equation*}
  \sum_s u(p,s)\bar u(p,s) = \pslash + m
  \eqsep,\eqsep
  \sum_s v(p,s)\bar v(p,s) = \pslash - m
  \eqsep.
\end{equation*}
The masses associated with the external momenta are specified in the second
argument by a rule or a list of rules. Here are some examples:
\begin{Verbatim}[commandchars=\\\{\}]
FermionPolarizationSums[Bar[USp[p]][s] . Ga[mu] . USp[p][s], p -> m]
\out DiracTr[(GI*m + Gs[p]) . Ga[mu]]
FermionPolarizationSums[Bar[USp[p]][s, al] USp[p][s, bt], p -> m]
\out (GI*m + Gs[p])[al, bt]
FermionPolarizationSums[
  Bar[USp[p]][r].VSp[k][s] Bar[VSp[k]][s].USp[p][r],
  {\lb}p -> m, k -> 0{\rb}]
\out DiracTr[(GI*m + Gs[p]) . Gs[k]]
\end{Verbatim}
Note that you can use patterns in the rules specifying the masses.
\begin{Verbatim}[commandchars=\\\{\}]
DeclareLorentzVectors[_q];
FermionPolarizationSums[
  Bar[USp[q[1]]][s] . Ga[mu] . USp[q[1]][s], q[n_] -> m[n]]
\out DiracTr[(Gs[q[1]] + GI*m[1]) . Ga[mu]]
\end{Verbatim}
If the mass associated with a certain external momentum is not specified the
corresponding spinors are left alone.
\begin{Verbatim}[commandchars=\\\{\}]
FermionPolarizationSums[
  Bar[USp[p]][r].VSp[k][s] Bar[VSp[k]][s].USp[p][r], k -> 0]
\out Bar[USp[p]][r] . Gs[k] . USp[p][r]
\end{Verbatim}

The function \code{VectorPolarizationSums} works in a similar way. It applies
the formulae
\begin{align*}
  \sum_s \epsilon^*_\mu(k,s)\epsilon_\nu(k,s) &=
    -g_{\mu\nu} + \frac{k_\mu k_\nu}{m^2}
    && \text{for massive particles,} \\
  \sum_s \epsilon^*_\mu(k,s)\epsilon_\nu(k,s) &=
    -g_{\mu\nu} - \frac{n^2 k_\mu k_\nu}{(n\cdot k)^2} +
                  \frac{n_\mu k_\nu + n_\nu k_\mu}{n\cdot k}
    && \text{for massless particles.}
\end{align*}
The polarisation sum for massless vector bosons depends on a \emph{gauge vector}
$n$ which must satisfy $n\cdot \epsilon(k,s)=0$ and $n\cdot k\neq 0$. The masses
of massive vector bosons and the gauge vectors of massless gauge bosons can be
passed to \code{VectorPolarizationSums} as a list of rules, similar to the
Fermion case:
\begin{Verbatim}[commandchars=\\\{\}]
VectorPolarizationSums[CC[MPol[k]][s, mu] MPol[k][s, nu], k -> m]
\out -Eta[mu, nu] + (k[mu]*k[nu])/m^2
DeclareLorentzVectors[n];
VectorPolarizationSums[CC[Pol[k]][s, mu] Pol[k][s, nu], k -> n]
\out -Eta[mu, nu] + (k[nu]*n[mu] + k[mu]*n[nu])/SDot[k, n] - 
  (k[mu]*k[nu]*Sqr[n])/SDot[k, n]^2
\end{Verbatim}
As in the case of \code{FermionPolarizationSums} you can use patterns in the
second argument and polarisation vectors are left alone if no mass or gauge
vector are specified for the external momentum.

You can apply an arbitrary function to the denominators of massless polarisation
sums with the \code{PostProcess} option. This is useful for performing
simplifications specific to your choice of the gauge vector. For example, if
you have chosen $n$ so that $n\cdot k=1$ you could implement this with
\begin{Verbatim}[commandchars=\\\{\}]
VectorPolarizationSums[CC[Pol[k]][s, mu] Pol[k][s, nu], k -> n,
  PostProcess -> ((# /. SDot[k, n] -> 1)&)]
\out -Eta[mu, nu] + k[nu]*n[mu] + k[mu]*n[nu] - k[mu]*k[nu]*Sqr[n]
\end{Verbatim}

%%%%%%%%%%%%%%%%%%%%%%%%%%%%%%%%%%%%%%%%%%%%%%%%%%%%%%%%%%%%%%%%%%%%%%%%%%%%%%%%
\subsection{One-Loop Integrals}
\label{sec:oneloop}
%%%%%%%%%%%%%%%%%%%%%%%%%%%%%%%%%%%%%%%%%%%%%%%%%%%%%%%%%%%%%%%%%%%%%%%%%%%%%%%%

A common technique in the computation of one-loop amplitudes is to express the
loop integrals in terms of \emph{Passarino-Veltman} (tensor) functions.  In
HEPMath this step is automated with the \code{PaVeIntegrate} function. Its
syntax is
\begin{Verbatim}[commandchars=\\\{\}]
PaVeIntegrate[\var{expr}, \var{l}, \var{muR}]
\end{Verbatim}
where \var{expr} is the expression to integrate, \var{l} is the loop momentum to
integrate over and \var{muR} is the renormalisation scale. This computes the
integral
\begin{equation*}
  (2\pi\var{muR})^{4-D}\int d^D\var{l}\ \var{expr}
\end{equation*}
where $D\equiv\code{Dim}$ is the dimension of Minkowski space. Propagator
denominators in \var{expr} must be represented by \code{Den} expressions (see
Tab.~\ref{tab:tens-predef}). \code{PaVeIntegrate} invokes
\code{PullVectors[expr, l]} and then replaces products of loop momenta and
propagator denominators with the correct linear combinations of
Passarino-Veltman one-loop tensor integrals. These integrals are represented
as follows
\begin{Verbatim}[commandchars=\\\{\}]
PaVe[\var{n},\var{muR},\var{irpow}][\var{indices}][\var{invariants}]
\end{Verbatim}
where \var{n} is the number of external legs (or propagator denominators) of the
integral, \var{muR} the renormalisation scale, \var{indices} the sequence of
indices identifying the tensor integral and \var{invariants} the sequence of
kinematic invariants specifying the external momenta and internal
masses. Currently only integrals with up to four propagator denominators are
supported. The argument \var{irpow} denotes the negative power of the infrared
regulator $\epsilon_\text{IR}$ in dimensional regularisation.  Its possible
values are 0, 1 and 2.

For example, the two-point rank two tensor integral can be computed as follows:
\begin{Verbatim}[commandchars=\\\{\}]
DeclareLorentzVectors[l, p];
DeclareReal[{m0, m1}];
PaVeIntegrate[l[mu] l[nu] Den[l, m0] Den[l+p, m1], l, muR]
\out I*Pi^2*(Eta[mu, nu]*(Div*((m0^2 + m1^2)/2 - Sqr[p]/12) + 
\noout    PaVe[2, muR, 0][0, 0][Sqr[p], m0^2, m1^2]) + 
\noout  p[mu]*p[nu]*(Div/3 + PaVe[2, muR, 0][1, 1][Sqr[p], m0^2, m1^2]))
\end{Verbatim}
The symbol \code{Div} represents the UV divergence
\begin{equation*}
  \code{Div} = \frac{2}{4 - \code{Dim}} - \gamma_\text{E} + \ln(4\pi)
\end{equation*}
where $\gamma_\text{E}$ is the Euler constant. The tensor integral
\begin{Verbatim}[commandchars=\\\{\}]
PaVe[2, muR, 0][1, 1][Sqr[p], m0^2, m1^2]
\end{Verbatim}
would usually be denoted as $B_{11}(p^2, m_0^2, m_1^2)$ with the dependence on
\code{muR} implicit. For the indices and order of arguments of the tensor
integrals HEPMath uses the same conventions as LoopTools, and I refer the
reader to the LoopTools manual for details.

By default \code{PaVeIntegrate} only returns the IR finite part of the integral.
If you set the option \code{IRDivergentParts} to \code{True} the result will be
given as a power series in \code{IRDiv}, which represents the IR divergence
$1/\epsilon_\text{IR}$.
\begin{Verbatim}[commandchars=\\\{\}]
PaVeIntegrate[Den[l, 0] Den[l+p, 0], l, muR, IRDivergentParts->True]
\out I*Pi^2*(Div + PaVe[2, muR, 0][0][Sqr[p], 0, 0] + 
\noout  IRDiv*PaVe[2, muR, 1][0][Sqr[p], 0, 0] + 
\noout  IRDiv^2*PaVe[2, muR, 2][0][Sqr[p], 0, 0])
\end{Verbatim}

%%%%%%%%%%%%%%%%%%%%%%%%%%%%%%%%%%%%%%%%%%%%%%%%%%%%%%%%%%%%%%%%%%%%%%%%%%%%%%%%
\subsection{Feynman Parametrisation of Multi-Loop Integrals}
\label{sec:multiloop}
%%%%%%%%%%%%%%%%%%%%%%%%%%%%%%%%%%%%%%%%%%%%%%%%%%%%%%%%%%%%%%%%%%%%%%%%%%%%%%%%

The standard method for the (analytic) computation of one-loop or multi-loop
integrals is to trade the integral over the loop momenta for an integral over
\emph{Feynman parameters}. In HEPMath you can do this transformation
automatically with the \code{FeynmanIntegrate} function. It only works for
\emph{scalar integrals}, i.e.\ for integrals without any loop momenta appearing
in the numerator. This means that before using \code{FeynmanIntegrate} you need
to cancel any appearences of loop momenta in the numerator against the
propagator denominators. The \code{DenCancel} function can help you with
that. It will cancel any \emph{squared} momenta in the numerator against
propagator denominators with the same momentum. However, it can not (yet)
eliminate scalar products between loop and external momenta or between different
loop momenta from the numerator. The syntax is
\begin{Verbatim}[commandchars=\\\{\}]
DenCancel[\var{expr}, \var{patt}]
\end{Verbatim}
This will cancel squares of momenta in \var{expr} which are not free of the
pattern \var{patt} against propagator denominators (\code{Den} expressions) with
the same momentum. If you omit the second argument it will try to cancel all
squared momenta. Note that the sign ambiguity of the arguments of \code{Sqr}
and \code{Den} is automatically taken care of:
\begin{Verbatim}[commandchars=\\\{\}]
DenCancel[Sqr[-l] Den[l, m], l]
\out 1 + m^2 Den[l, m]
\end{Verbatim}

The syntax of \code{FeynmanIntegrate} is
\begin{Verbatim}[commandchars=\\\{\}]
FeynmanIntegrate[\var{expr}, {\lb}\var{l1}, \ldots{}, \var{lL}{\rb}, \var{x}, \var{muR}, \var{del}]
\end{Verbatim}
where \var{expr} is the integrand expression, \code{\{\var{l1}, \ldots{},
  \var{lL}\}} is the list of loop momenta, \var{muR} is the renormalisation
scale, and \var{del} is the symbol used for the infinitesimal imaginary part of
the propagators. The argument \var{x} is the head used to generate the Feynman
parameters \code{\var{x}[1]}, \code{\var{x}[2]}, etc. You can omit the \var{del}
argument, in which case it is set to zero. \code{FeynmanIntegrate} computes the
integral
\begin{equation*}
  (2\pi\var{muR})^{L(4-\code{Dim})} \int d^{\code{Dim}}\var{l1}\cdots
  \int d^{\code{Dim}}\var{lL}\ \var{expr}
  \eqsep,
\end{equation*}
where $L$ is the number of loop momenta.  The resulting Feynnman integrals are
of the form
\begin{equation}
  \mathcal{G}(D)\int_0^1 dx_1\cdots\int_0^1 dx_n \delta(1-x_1-\ldots-x_n)
  \mathcal{M}(\vec x)\mathcal{F}(\vec x)^{\alpha(D)}
  \mathcal{U}(\vec x)^{\beta(D)}
  \eqsep,
\end{equation}
where $\mathcal{G}(D)$ is a pre-factor involving Gamma functions and the
space-time dimension $D(\equiv\code{Dim})$, $\mathcal{M}(\vec x)$ is a monomial
in the Feynman parameters $x_i$ originating from denominators raised to some
power, $\mathcal{F}(\vec x)$ and $\mathcal{U}(\vec x)$ are polynomials in the
$x_i$ and $\alpha(D)$ and $\beta(D)$ are powers depending on $D$. In HEPMath
these integrals are represented by \code{FeynmanIntegral} expressions, which
have the following form:
\begin{Verbatim}[commandchars=\\\{\}, codes={\catcode`$=3\catcode`^=7\catcode`_=8}]
FeynmanIntegral[{\lb}$\mathcal{G}(D)$, $\mathcal{M}(\vec{x})$, $\mathcal{F}(\vec{x})^{\alpha(D)}$, $\mathcal{U}(\vec{x})^{\beta(D)}${\rb},
                {\lb}$x_1$, \ldots{}, $x_n${\rb}]
\end{Verbatim}
The first thing you usually want to do with these integrals is integrate out the
delta function by eliminating one Feynman parameter. (Since you may want to
have a say in which Feynman parameter gets eliminated HEPMath doesn't do this
automatically.) This can be done with the \code{EliminateFeynmanParameter}
function:
\begin{Verbatim}[commandchars=\\\{\}]
EliminateFeynmanParameter[\var{f}, x[\var{i}]]
\end{Verbatim}
integrates the delta function by eliminating the \var{i}-th Feynman parameter.
If you drop the second argument the parameter is chosen automatically. The
argument \var{f} must be a \code{FeynmanIntegral} expression, so you'll typically
use a replacement rule like
\begin{Verbatim}[commandchars=\\\{\}]
expr = expr /. f_FeynmanIntegral :> EliminateFeynmanParameter[f]
\end{Verbatim}
The expressions returned by \code{EliminateFeynmanParameter} still contain
integrals over the remaining Feynman parameters. These integrals are represented
by \code{HEPIntegral} expressions. \code{HEPIntegral} has the same syntax as
Mathematica's \code{Integrate} function, but it does not attempt to
compute the integrals. It is just a container for intermediate expressions
which you will need to manipulate further to compute the integrals. However,
you can expand \code{HEPIntegral} expressions with \code{HEPExpand}.

%%%%%%%%%%%%%%%%%%%%%%%%%%%%%%%%%%%%%%%%%%%%%%%%%%%%%%%%%%%%%%%%%%%%%%%%%%%%%%%%
\section{User-Defined Tensors and Linear Functions}
%%%%%%%%%%%%%%%%%%%%%%%%%%%%%%%%%%%%%%%%%%%%%%%%%%%%%%%%%%%%%%%%%%%%%%%%%%%%%%%%

Pre-defined tensors such as \code{Ga} or \code{Eps} are actually just examples
of HEPMath's capability of integrating arbitrary tensors and index types in
Mathematica. To see how this works let us try to add the notion of Weyl spinors
to HEPMath. Note that the code in this section is just an example.  Weyl
spinors are (currently) \emph{not} included in HEPMath.

First we have to add an index type which represents Weyl spinors. (For
simplicity we will not distinguish between dotted, undotted, raised and lowered
indices). New index types can be declared with the \code{DeclareIndexType}
function:
\begin{Verbatim}[commandchars=\\\{\}]
DeclareIndexType[Weyl, 2, WI]
\end{Verbatim}
This introduces a new index type called \code{Weyl} with index dimension 2. It
also tells HEPMath that the identity matrix in the space of Weyl spinors is
represented by \code{WI}. Note that \code{HEPContract} immediately knows what
to do with \code{WI}:
\begin{Verbatim}[commandchars=\\\{\}]
HEPContract[WI[aa, bb] WI[bb, cc]]
\out WI[aa, cc]
HEPContract[WI[aa, aa]]
\out 2
\end{Verbatim}

Next we introduce the $\sigma^\mu$ and $\bar\sigma^\mu$ matrices and the
anti-symmetric tensor in Weyl-space:
\begin{Verbatim}[commandchars=\\\{\}]
DeclareHEPTensor[{\lb}WSig, WSigBar{\rb}, {\lb}Lorentz, Weyl, Weyl{\rb}]
DeclareHEPTensor[WEps, {\lb}Weyl, Weyl{\rb}]
\end{Verbatim}
Functions like \code{HEPTensorSignature}, \code{Indices}, \code{ExplicitIndices}
or \code{HEPExpand} now accept the new objects as first-class citizens.
\begin{Verbatim}[commandchars=\\\{\}]
DeclareLorentzVectors[p];
HEPTensorSignature[p.WSig]
\out {\lb}Weyl, Weyl{\rb}
Indices[WSig[mu, aa, bb] WSigBar[mu, bb, cc]]
\out {\lb}{\lb}aa -> Weyl, cc -> Weyl{\rb}, {\lb}bb -> Weyl, mu -> Lorentz{\rb}{\rb}
ExplicitIndices[(p.WSig)[aa,bb]]
\out p[$3]*WSig[$3, aa, bb]
HEPExpand[(p.(a WSig + b WSigBar))[aa,bb]]
\out a*(p . WSig)[aa, bb] + b*(p . WSigBar)[aa, bb]
\end{Verbatim}
We may also want to tell HEPMath that \code{WEps} is real.
\begin{Verbatim}[commandchars=\\\{\}]
DeclareReal[WEps];
CC[WEps[aa, bb]]
\out WEps[aa, bb]
\end{Verbatim}
Note that there is currently no general way to declare a tensor as
anti-symmetric. To exploit this property you will have to write your own
simplification function.

Traces over Dirac and colour indices are represented by \code{DiracTr} and
\code{ColorTr}, respectively, and we may want to have a similar notation for
Weyl indices.  We therefore introduce a new symbol \code{WeylTr} which
represents a trace over Weyl indices. There are two properties of \code{WeylTr}
which HEPMath needs to know about: the trace is \emph{linear} and
\emph{real holomorphic}, meaning that the complex conjugate of a trace is the
same as the trace of the complex conjugate of its argument. These two properties 
can be declared separately with
\begin{Verbatim}[commandchars=\\\{\}]
DeclareLinear[WeylTr];
DeclareRealHolomorphic[WeylTr];
HEPExpand[WeylTr[(a WSig + b WSigBar)[mu]]]
\out a*WeylTr[WSig[mu]] + b*WeylTr[WSigBar[mu]]
CC[WeylTr[WSig[mu] + p[mu] WEps]]
\out WeylTr[WEps*p[mu] + CC[WSig][mu]]
\end{Verbatim}
Note that declaring a function as linear means that it is linear in \emph{all
  its arguments}.  \code{DeclareLinear[WeylTr]} not only instructs
\code{HEPExpand} how to operate on \code{WeylTr} expressions, it also tells
HEPMath that \code{WeylTr} ``exports'' any free or contracted indices appearing
in its argument:
\begin{Verbatim}[commandchars=\\\{\}]
Indices[WeylTr[p[mu] WSig[mu].WSigBar[nu]]]
\out {\lb}{\lb}nu -> Lorentz{\rb}, {\lb}mu -> Lorentz{\rb}{\rb}
\end{Verbatim}
If a function is not declared as linear HEPMath will simply ignore any 
indices appearing in its arguments:
\begin{Verbatim}[commandchars=\\\{\}]
Indices[somefunc[p[mu] WSig[mu].WSigBar[nu]]]
\out {\lb}{\lb}{\rb}, {\lb}{\rb}{\rb}
\end{Verbatim}

%%%%%%%%%%%%%%%%%%%%%%%%%%%%%%%%%%%%%%%%%%%%%%%%%%%%%%%%%%%%%%%%%%%%%%%%%%%%%%%%
\section{Interfaces}
%%%%%%%%%%%%%%%%%%%%%%%%%%%%%%%%%%%%%%%%%%%%%%%%%%%%%%%%%%%%%%%%%%%%%%%%%%%%%%%%

In cases where many different tools are used in the same project the development
of interfaces between these tools can take a substantial amount of time. With
its consistent implementation of High Energy Physics notations in Mathematica,
HEPMath is the ideal environment for combining different tools related to High
Energy Physics. HEPMath currently provides interfaces to FeynArts, LoopTools
and LHAPDF, but I hope that this list will get longer in the future.  This
section describes the interfaces currently present in HEPMath.

%%%%%%%%%%%%%%%%%%%%%%%%%%%%%%%%%%%%%%%%%%%%%%%%%%%%%%%%%%%%%%%%%%%%%%%%%%%%%%%%
\subsection{FeynArts}
%%%%%%%%%%%%%%%%%%%%%%%%%%%%%%%%%%%%%%%%%%%%%%%%%%%%%%%%%%%%%%%%%%%%%%%%%%%%%%%%

The FeynArts interface of HEPMath (obviously) requires a working FeynArts
installation. HEPMath was tested with FeynArts version 3.8. To use FeynArts
from within HEPMath you must \emph{not} load the FeynArts context directly.
Instead you call
\begin{Verbatim}[commandchars=\\\{\}]
Needs["HEPMath`FeynArts`"]
\end{Verbatim}
This will give you access to all FeynArts functions and symbols, but it moves
some of the FeynArts symbols to a different context in order to avoid name
clashes with HEPMath.

After loading the FeynArts interface you can then use FeynArts in the same way
as described in the FeynArts manual, except for one small difference: if you
need access to symbols defined in your FeynArts model file you \emph{must}
initialise the model before these symbols are mentioned. The reason is that
HEPMath tweaks FeynArts\footnote{Not permanently. The FeynArts source files are
  not modified and loading FeynArts without HEPMath will give you the usual
  behaviour.} so that it does not load models into the global context but in the
context \code{HEPMath`FeynArts`Model} instead. This way the global context does
not get polluted and name clashes between symbols in the model file and HEPMath
symbols can be avoided. To make sure that all symbols from the model file end up
in \code{HEPMath`FeynArts`Model} and not \code{Global`} you have to initialise
the model before using any of its symbols. Here is an example of what can go
wrong:
\begin{Verbatim}[commandchars=\\\{\}]
Needs["HEPMath`"];
Needs["HEPMath`FeynArts`"];
(* InitializeModel["SM"]; *)
SetOptions[InsertFields, Model -> "SM",
           InsertionLevel -> {\lb}Particles{\rb},
           Restrictions -> NoLightFHCoupling];
process = {\lb}S[1]{\rb} -> {\lb}F[4, {\lb}3{\rb}], -F[4, {\lb}3{\rb}]{\rb};
topos = CreateTopologies[0, 1 -> 2];
ins = InsertFields[tops, process];
amp = CreateFeynAmp[ins];
\out InitializeModel::badrestr: 
\noout   Warning: Global`NoLightFHCoupling is not a valid model
\noout   restriction.
\end{Verbatim}
Removing the comments around \code{InitializeModel["SM"];} gets rid of the
warning and everything will run smoothly.

To manipulate the amplitude \code{amp} from the example above in HEPMath you
first have to convert it to HEPMath notation. For the model files included in
the FeynArts distribution this can be done with the \code{ConvertFeynAmp}
function. Continuing the example above you can call
\begin{Verbatim}[commandchars=\\\{\}]
{\lb}cfg, amp{\rb} = ConvertFeynAmp[amp];
cfg
\out {\lb}Mass[Incoming, 1] -> MH, Mass[Outgoing, 1] -> MT,
\noout  Mass[Outgoing, 2] -> MT, Indices[Incoming, 1] -> {\lb}{\rb},
\noout  Indices[Outgoing, 1] -> {\lb}Index[Colour, 2]{\rb}, 
\noout  Indices[Outgoing, 2] -> {\lb}Index[Colour, 3]{\rb}{\rb}
amp
\out I*ColorDelta[Index[Colour, 2], Index[Colour, 3]]*
\noout  Bar[USp[FourMomentum[Outgoing, 1]]][Polarization[Outgoing, 1]] . 
\noout   ((-I/2*EL*MT*PL)/(MW*SW) - (I/2*EL*MT*PR)/(MW*SW)) . 
\noout   VSp[FourMomentum[Outgoing, 2]][Polarization[Outgoing, 2]]
\end{Verbatim}
\code{cfg} is a list which holds information about the masses and indices of the
external particles and \code{amp} is an expression which represents the Feynman
amplitude. \code{Polarization[Outgoing, 1]} denotes the spin polarisation index
of the first outgoing particle (i.e. the top quark). Note that the
\code{FourMomentum} expressions are recognised as Lorentz vectors by HEPMath.
\begin{Verbatim}[commandchars=\\\{\}]
HEPTensorSignature[FourMomentum[Outgoing, 1]]
\out {\lb}Lorentz{\rb}
\end{Verbatim}
Thus the expression \code{amp} can be directly manipulated with the HEPMath
functions described in Sec.~\ref{sec:algebra}. To make intermediate results
easier to read I usually introduce some shorthands like
\begin{Verbatim}[commandchars=\\\{\}]
DeclareLorentzVectors[_k, _p];
amp = amp /. {\lb}FourMomentum[Incoming, n_] -> k[n],
              FourMomentum[Outgoing, n_] -> p[n],
              FourMomentum[Internal, n_] -> l[n],
              Polarization[Incoming, n_] -> sg[n],
              Polarization[Outgoing, n_] -> lm[n],
              Index[Gluon, n_]           -> a[n],
              Index[Colour, n_]          -> i[n],
              Index[Lorentz, n_]         -> mu[n]{\rb}
\out I*ColorDelta[i[2], i[3]]*Bar[USp[p[1]]][lm[1]] . 
\noout   ((-I/2*EL*MT*PL)/(MW*SW) - (I/2*EL*MT*PR)/(MW*SW)) .
\noout   VSp[p[2]][lm[2]]
\end{Verbatim}
but this is purely a matter of taste.

The function \code{ConvertFeynAmp} should work for amplitudes generated with one
of the model files shipping with FeynArts. However, since FeynArts does not
really define the set of kinematic objects which are allowed to appear in
model files (and thus in expressions returned by \code{CreateFeynAmp}) it
is not possible to write a conversion function that is guaranteed to work with
any model file. If you have a model file that is not supported by HEPMath and
you think it should be feel free to let me know.

%%%%%%%%%%%%%%%%%%%%%%%%%%%%%%%%%%%%%%%%%%%%%%%%%%%%%%%%%%%%%%%%%%%%%%%%%%%%%%%%
\subsection{LoopTools}
%%%%%%%%%%%%%%%%%%%%%%%%%%%%%%%%%%%%%%%%%%%%%%%%%%%%%%%%%%%%%%%%%%%%%%%%%%%%%%%%

The LoopTools package \cite{hep-ph/9807565} allows you to evaluate one-loop
tensor integrals numerically. If you enabled the LoopTools interface during
the installation you can load it in your Mathematica session with
\begin{Verbatim}[commandchars=\\\{\}]
Needs["HEPMath`LoopTools`"]
\end{Verbatim}
The \code{PaVe} expressions introduced in Sec.~\ref{sec:oneloop} will now 
evaluate to numbers when you call them with numeric arguments.
\begin{Verbatim}[commandchars=\\\{\}]
PaVe[2,1.0,0][0][10.0, 1.0, 2.0]
\out 0.622592 + 2.0116 I
\end{Verbatim}
If you want to generate numerical code which calls functions from LoopTools
you also need the LoopTools interface installed.

%%%%%%%%%%%%%%%%%%%%%%%%%%%%%%%%%%%%%%%%%%%%%%%%%%%%%%%%%%%%%%%%%%%%%%%%%%%%%%%%
\subsection{LHAPDF}
%%%%%%%%%%%%%%%%%%%%%%%%%%%%%%%%%%%%%%%%%%%%%%%%%%%%%%%%%%%%%%%%%%%%%%%%%%%%%%%%

The LHAPDF library is a common C++ interface for a large number of parton
distribution functions (PDFs). HEPMath provides a simple interface to LHAPDF
which lets you load PDF sets and evaluate PDFs numerically in Mathematica.  As
in the case of LoopTools you need to enable the LHAPDF interface during
installation to use it in Mathematica, and also to generate code which calls
LHAPDF functions.

You can load the interface with
\begin{Verbatim}[commandchars=\\\{\}]
Needs["HEPMath`LHAPDF`"]
\end{Verbatim}
To load a specific PDF set you call, for example,
\begin{Verbatim}[commandchars=\\\{\}]
pdfid = LHAPDFOpen["cteq6l1", 0]
\end{Verbatim}
The first argument identifies the collection of PDFs you want to use and the
second argument is the number of the desired PDF set within that collection.
The function loads the PDF set into Mathematica and returns the ``LHAPDF-ID'',
a number which uniquely identifies that set. If you know that number
you can also load the set with
\begin{Verbatim}[commandchars=\\\{\}]
LHAPDFOpenID[\var{pdfid}]
\end{Verbatim}
To get the number of PDF sets in a given collection (and thus the allowed range
of values for the second argument of \code{LHAPDFOpen}) you can call
\begin{Verbatim}[commandchars=\\\{\}]
LHAPDFMembers["cteq6l1"]
\out 1
\end{Verbatim}
The \code{cteq6l1} collection only contains one set.

Once you have loaded the PDF set you want and stored its ID in \code{pdfid}
you can evaluate the PDF with
\begin{Verbatim}[commandchars=\\\{\}]
LHAPDF[pdfid, \var{pid}, \var{x}, \var{Q}]
\end{Verbatim}
Here \var{pid} is the PDG code of the parton, \var{x} is the momentum fraction
and \var{Q} the factorisation scale. To get a list of PDG codes of the partons
included in the PDF set call
\begin{Verbatim}[commandchars=\\\{\}]
LHAPDFFlavors[pdfid]
\end{Verbatim}
To compute the value of $\alpha_s$ associated with the PDF set at the scale
\var{Q} use
\begin{Verbatim}[commandchars=\\\{\}]
LHAPDFAlphaS[pdfid, \var{Q}]
\end{Verbatim}
When you are done with the PDF set remember to close it with
\begin{Verbatim}[commandchars=\\\{\}]
LHAPDFClose[pdfid]
\end{Verbatim}

%%%%%%%%%%%%%%%%%%%%%%%%%%%%%%%%%%%%%%%%%%%%%%%%%%%%%%%%%%%%%%%%%%%%%%%%%%%%%%%%
\section{Code Generation}
%%%%%%%%%%%%%%%%%%%%%%%%%%%%%%%%%%%%%%%%%%%%%%%%%%%%%%%%%%%%%%%%%%%%%%%%%%%%%%%%

More often than not the result of an analytical calculation can not be
simplified to a digestible form, and typically some extensive numerical
procedure needs to be applied to the result. For cross sections this procedure
is usually the phase space integration. In other cases one might want to include
the result in a fit where the computed quantity contributes to a $\chi^2$
function which is minimised numerically.

Of course Mathematica is perfectly capable of doing numerical calculations.  The
efficiency of these calculations can be significantly increased with
Mathematica's \code{Compile} function, which compiles Mathematica expressions
into byte-code. With the recently added support for native compilation targets
the performance is actually comparable to hand-written C code. However, due to
the limited availability of Mathematica licenses at most institutes one needs a
different solution when the numerical computation is to be parallelised and run
on a computer cluster.

The conventional approach is to use Mathematica's \code{CForm} or
\code{FortranForm} functions to convert expressions into C or Fortran code
snippets which are then pieced together and compiled outside Mathematica.  The
disadvantages of this approach are that it is not very general and that any code
optimisation which is beyond the compiler's capabilities must be implemented by
the user (or by the package providing the code generation functionality, as in
the case of FormCalc).  Since Mathematica version 8 there is actually a viable
alternative. If the function only uses a certain subset of \emph{compilable}
Mathematica functions the byte-code generated by Mathematica's \code{Compile}
function can now be exported to C code. \code{Compile} does a rather decent job
at optimising large expressions, and since the C code is generated from
byte-code rather than Mathematica expressions it benefits from the optimisation
done by \code{Compile}. This method of generating code is not only efficient but
also very flexible.  You can first use Mathematica to write, test and compile a
function which does exactly what you want, and then put the compiled function
into a shared library which you can link with the rest of your code. The code
generation system implemented in HEPMath is based on this strategy.

Mathematica's code generation from byte-code currently has a major shortcoming:
it is not possible for the generated code to call functions from other
libraries. HEPMath provides a workaround for this by post-processing the
generated code. As a result the code generated by HEPMath can call any external
function which is interfaced to Mathematica via the \code{LibraryLink}
interface.

HEPMath provides a function \code{HEPCompile} which mirrors the behaviour of
Mathematica's \code{Compile} function but guarantees the correct treatment of
symbols introduced by HEPMath. The use of \code{HEPCompile} is best explained
by examples:
\begin{Verbatim}[commandchars=\\\{\}]
Needs["HEPMath`Compile`"];
expr = x^2 + y^2;
cf1 = HEPCompile[{\lb}{\lb}x, _Real{\rb}, {\lb}y, _Real{\rb}{\rb},
  HEPEvaluate[expr] + 1.0];
cf1[1.0, 1.0]
\out 3.
\end{Verbatim}
The first argument to \code{HEPCompile} specifies the names and types of
arguments to the function. Its syntax is identical to the one used in
\code{Compile}. The second argument is the function body. Like \code{Compile},
\code{HEPCompile} does not evaluate the function body. (It has the attribute
\code{HoldAll}.) However, since you usually want to use \code{HEPCompile} to
compile large expressions and you probably don't want to type them in by hand
you can selectively evaluate parts of the function body with \code{HEPEvaluate}.
Wrapping \code{HEPEvaluate} around an expression inside a \code{HEPCompile}
function body instructs \code{HEPCompile} to substitute the evaluation of that
expression before compilation. The body of the above function is therefore
\code{x\^{}2 + y\^{}2 + 1.0} and thus evaluates to 3.0 when called with
\code{x=y=1.0}.

\code{HEPCompile} can also handle function bodies that depend on Lorentz vectors
and contain one-loop integrals. In the function body of \code{HEPCompile}
Lorentz vectors are represented by lists of four numbers (the first element
being the time component). However, before injecting symbolic expressions which
contain Lorentz vectors or \code{PaVe} expressions in a function body some
additional manipulations are necessary. You can let \code{HEPCompile} do these
manipulations automatically by wrapping \code{HEPPrepare} around the relevant
parts of the function body:
\begin{Verbatim}[commandchars=\\\{\}]
DeclareLorentzVectors[p];
b0 = PaVe[2, 1.0, 0][0][Sqr[p], m^2, m^2];
cf2 = HEPCompile[{\lb}{\lb}p, _Real, 1{\rb}, {\lb}m, _Real{\rb}{\rb},
  HEPPrepare[b0]];
cf2[{\lb}5.0, 0.0, 0.0, 0.0{\rb}, 2.0]
\out -0.218071 + 1.88496 I
\end{Verbatim}

In addition to \code{HEPEvaluate}, \code{HEPPrepare} and the compilable
functions supported by \code{Compile} (call Mathematica's undocumented function
\code{Compile`CompilerFunctions[]} for a list) you can use the following
expressions inside a \code{HEPCompile} body:
\begin{itemize}
\item \code{Pol[\var{p}][\var{s}]} and \code{MPol[\var{p}][\var{s}]}
  return the polarisation vectors of a massless and massive vector boson with
  momentum \var{p} and helicity \var{s}. The argument \var{p} and the return
  values are lists of four numbers and \var{s} must be an integer
  which can be -1 or 1 for \code{Pol} and -1, 0 or 1 for \code{MPol}.
\item \code{LHAPDFOpenID}, \code{LHAPDF}, \code{LHAPDFAlphaS} and
  \code{LHAPDFClose} with their correct argument lists.
\item \code{KineticLambda[\var{x},\var{y},\var{z}]} represents the kinetic
  function $\lambda(x,y,z)=x^2 + y^2 + z^2 - 2(xy + zy + zx)$.
\item \code{TwoBodyDecay[\var{q}, \var{m1}, \var{m2}, \var{theta}, \var{phi}]}
  can be used for constructing phase space parametrisations. It returns a pair
  of Lorentz vectors (i.e. lists of four numbers) which are the momenta of the
  decay products of a particle with four momentum \var{q} decaying into two
  particles with masses \var{m1} and \var{m2}. The arguments \var{theta} and
  \var{phi} are polar and azimuthal angles in the rest frame of the decaying
  particle.
\end{itemize}

Once you have compiled all the functions you need you can export them with the
\code{HEPCodeGenerate} function. It mirrors the behaviour of Mathematica's
\code{CCodeGenerate} function but should only be used on compiled functions
generated with \code{HEPCompile}. To export the functions \code{cf1} and
\code{cf2} from the examples above to a library called \code{myfuncs} execute
\begin{Verbatim}[commandchars=\\\{\}]
modules = HEPCodeGenerate[{\lb}cf1, cf2{\rb}, {\lb}"f1", "f2"{\rb}, "myfuncs",
  TargetDirectory->"build"]
\out {\lb}"build/myfuncs.c", "build/myfuncs_f1.c", "build/myfuncs_f2.c"{\rb}
\end{Verbatim}
This puts a number of files in the directory \code{build} and returns a list
with the names of the C source files it generated. Make sure the directory
\code{build} exists before you run this. You will also find a file called
\code{myfuncs.py} in that directory. This is the python wrapper which will
allow you to call your functions from Python. But first you have to compile
the sources into a shared library. You can do this directly from Mathematica
by using the \code{CCompilerDriver} package:
\begin{Verbatim}[commandchars=\\\{\}]
Needs["CCompilerDriver`"];
CreateLibrary[modules, "_myfuncs",
  "TargetDirectory" -> "build",
  "Debug"->True, "Libraries"->{\lb}"WolframRTL", "dl"{\rb},
  "ShellCommandFunction"->Print]
\end{Verbatim}
The advantage of this method is that all the flags needed to compile against the
Wolfram run-time library will be set automatically by Mathematica. You can also
do this manually, e.g.\ in a makefile. In that case simply copy the flags from
the output generated by the \code{CreateLibrary} call above. Note that the first
argument to \code{CreateLibrary} is the list of modules returned by
\code{HEPCodeGenerate}. The second argument is the name of your library (without
the extension). If you want to use the Python interface this should always be
the name specified in \code{HEPCodeGenerate} preceded by an underscore. Also note
that you have to specify the Wolfram run-time library and the \code{dl} library
in the \code{"Libraries"} option. The latter is needed for external calls to
other libraries, e.g. to LoopTools.

If everything worked you should now have a file \code{\_myfuncs.so} in your
\code{build} directory. To use this library you first have to make sure that
the Wolfram run-time library is found by your linker. This can be accomplished by
adding the location of \code{libWolframRTL.so} to your \code{LD\_LIBRARY\_PATH}
variable. On my system this location is
\begin{Verbatim}[commandchars=\\\{\}]
/usr/local/Wolfram/Mathematica/9.0/SystemFiles/Libraries/Linux-x86-64
\end{Verbatim}
but it might be somewhere else on yours. When this is done you can go into
the \code{build} directory, open a Python session and run
\begin{Verbatim}[commandchars=\\\{\}]
>>> import myfuncs
>>> myfuncs.f1(1.0, 1.0)
3.0
>>> myfuncs.f2([5.0, 0.0, 0.0, 0.0], 2.0)
(-0.21807097779182483+1.8849555921538759j)
\end{Verbatim}
The Lorentz vectors can be passed as lists or as \code{numpy} arrays. If your
function would return a list or a matrix in Mathematica this will also become
a \code{numpy} array in Python.

To have the \code{myfuncs} module availble in all your Python sessions just move
the files \code{myfuncs.py} and \code{\_myfuncs.so} to some place where Python
will find them (e.g. a directory listed in your \code{PYTHONPATH} variable).
Just make sure that the wrapper module (\code{myfuncs.py}) and the
library (\code{\_myfuncs.so}) are always kept in the same directory.

%%%%%%%%%%%%%%%%%%%%%%%%%%%%%%%%%%%%%%%%%%%%%%%%%%%%%%%%%%%%%%%%%%%%%%%%%%%%%%%%
\section{Performance}
\label{sec:performance}
%%%%%%%%%%%%%%%%%%%%%%%%%%%%%%%%%%%%%%%%%%%%%%%%%%%%%%%%%%%%%%%%%%%%%%%%%%%%%%%%

HEPMath is mainly a convenience tool with a very broad range of
applications. Its performance (in terms of speed) was not a major concern during
its development and can not be quantified with a single or even a few benchmark
calculations. However, to give you at least a rough idea of its speed I use the
analytical computation and the numerical evaluation of the squared matrix
element for the Standard Model process $gg \to Hg$ as benchmarks. The
corresponding Mathematica script can be found under \code{examples/gg-gH-SM.m}
in the source distribution. I compare the performance of HEPMath with FormCalc
\cite{hep-ph/9807565, hep-ph/0210220, hep-ph/0406288, hep-ph/0506201,
  hep-ph/0601248, hep-ph/0607049, hep-ph/0611273} and MadLoop
\cite{arXiv:1103.0621}. The results are summarised in
Tab.~\ref{tab:performance}. The second column holds the real time (in seconds)
needed for the analytical computations, code generation and compilation. The
third column holds the real time (in milliseconds) needed for one evaluation of
the squared matrix element. This was obtained by averaging 1000 evaluations.
The tests were run on an
Intel\textsuperscript{\textregistered}\ Core\textsuperscript{TM} i5 CPU 760 at
\unit{2.8}{GHz}.

\ctable[
  caption={Comparison of the performances of HEPMath, FormCalc
    \cite{hep-ph/9807565, hep-ph/0210220, hep-ph/0406288, hep-ph/0506201,
      hep-ph/0601248, hep-ph/0607049, hep-ph/0611273}, and MadLoop
    \cite{arXiv:1103.0621} for the SM process $gg\to Hg$ on an
    Intel\textsuperscript{\textregistered}\ Core\textsuperscript{TM} i5 CPU 760
    at \unit{2.8}{GHz}.},
  label=tab:performance
]{lrr}{}{
  \FL
  tool     &  code generation & matrix element evaluation \ML
  FormCalc &  \unit{4.7}{sec} &          \unit{1.8}{msec} \NN
  MadLoop  & \unit{10.2}{sec} &         \unit{24.4}{msec} \NN
  HEPMath  & \unit{34.3}{sec} &         \unit{17.8}{msec} \LL
}

We see that FormCalc beats both tools in both stages of the computation.
FormCalc's matrix element evaluations (at least for this process) are faster by
an order of magnitude.  The performace of HEPMath is similar to that of
MadLoop. HEPMath's code generation procedure is slower than that
of MadLoop by about a factor of 3, but the numerical evaluation of the squared
matrix element is slightly faster.

%%%%%%%%%%%%%%%%%%%%%%%%%%%%%%%%%%%%%%%%%%%%%%%%%%%%%%%%%%%%%%%%%%%%%%%%%%%%%%%%
\section{Conclusions}
%%%%%%%%%%%%%%%%%%%%%%%%%%%%%%%%%%%%%%%%%%%%%%%%%%%%%%%%%%%%%%%%%%%%%%%%%%%%%%%%

This concludes our tour of the Mathematica package \emph{HEPMath}. Its goal is
not to implement a particular computational algorithm or set of algorithms but
to make the implementation of symbolic computations related to High Energy
Physics easier. It accomplishes this with two features which set it apart from
similar public codes: a consistent and extensible integration of tensors and
index contractions in the Mathematica language and a flexible mechanism for
generating numerical code which allows the direct transition between two
high-level programming languages: Mathematica and Python. In addition it
provides seamless interfaces to three popular tools: FeynArts, LoopTools and
LHAPDF. It still lacks some of the more advanced features present in other
packages like the Passarino-Veltman reduction to scalar integrals or the
automatic computation of renormalisation constants. I hope to remedy this
situation in future releases.  HEPMath was recently used for a computation of
Higgs+jet cross sections in the presence of effective dimension six operators
\cite{arXiv:1411.2029}.

\subsubsection*{Acknowledgements}
I would like to thank Ulrich Nierste, Stefan Schacht and Wolfgang Noll for
helpful comments and beta-testing, and Olivier Mattelaer for help with the
performance comparison with MadLoop.

\bibliography{references}
\end{document}